%% file: ms.tex
\documentclass[a4paper,11pt]{article}
\pdfoutput=1 
\usepackage{jcappub}
\usepackage{graphicx}
\pdfoutput=1 
\usepackage{color}
\usepackage[dvipsnames]{xcolor}
\usepackage{multirow}
\usepackage{cleveref}
\usepackage[normalem]{ulem}		
\usepackage[utf8]{inputenc}

\newcommand{\rr}{\mathrm}

\title{The promising future of a robust cosmological neutrino mass measurement}

\author[1]{Thejs Brinckmann,}
\author[1]{Deanna C. Hooper,}
\author[1]{Maria Archidiacono,}
\author[1]{Julien Lesgourgues,}
\author[1]{and Tim Sprenger}

\affiliation[1]{Institute for Theoretical Particle Physics and Cosmology (TTK), \\ RWTH Aachen University, D-52056 Aachen, Germany.}

\emailAdd{brinckmann@physik.rwth-aachen.de}
\emailAdd{hooper@physik.rwth-aachen.de}

\abstract{We forecast the sensitivity of thirty-five different combinations of future Cosmic Microwave Background and Large Scale Structure data sets to cosmological parameters and to the total neutrino mass. We work under conservative assumptions accounting for uncertainties in the modelling of systematics. In particular, for galaxy redshift surveys, we remove the information coming from non-linear scales. We use Bayesian parameter extraction from mock likelihoods to avoid Fisher matrix uncertainties. Our grid of results allows for a direct comparison between the sensitivity of different data sets. We find that future surveys will measure the neutrino mass with high significance and will not be substantially affected by potential parameter degeneracies between neutrino masses, the density of relativistic relics, and a possible time-varying equation of state of Dark Energy.}

\begin{document}

\hfill{\small TTK-18-29}

\maketitle

\input{introduction.tex}
\input{models.tex}
\input{experiments.tex}
\input{results.tex}
\input{conclusions.tex}

\section*{Acknowledgements}
We thank Jacques Delabrouille for very useful exchanges. Simulations for this work were performed with computing resources granted by the RWTH High Performance Computing cluster under project rwth0113 and by JARA-HPC from RWTH Aachen University under jara0184.

\clearpage
\appendix
\input{appendix.tex}
\clearpage

\bibliography{references}{}
\bibliographystyle{JHEP}

\end{document}

%% file: introduction.tex
\section{Introduction}

Cosmological observables are known to be very sensitive to the sum of neutrino masses \cite{Hu:1997mj,Lesgourgues:2006nd,Lesgourgues:1519137,Archidiacono:2016lnv,Lattanzi:2017ubx}, with the potential to detect the neutrino mass scales well before laboratory experiments. This conclusion is often tempered by the fact that cosmological bounds depend on underlying assumptions on the cosmological model, since all bounds are derived from global fits to the observed data set. Fortunately, different ingredients in the cosmological model usually have distinct effects, and the cosmological data sets result in thousands of independent data points, such that in many cases parameter degeneracies can be broken. Nonetheless, bounds on parameters like the total neutrino mass can weaken significantly when more complicated cosmological models with extra free parameters are considered. Over the next decades, we expect increasingly precise data on the power spectrum of Cosmic Microwave Background (CMB) anisotropies and Large Scale Structure (LSS). This will allow us to get not just stronger bounds on the total neutrino mass, but also more robust and model-independent bounds.

Therefore, a very interesting question to address is: which combination of datasets do we need in order to detect the total neutrino mass at a given confidence level, not just assuming a minimal underlying cosmological model, but also extended models? And, at which point will we be able to resolve parameter degeneracies?  

These questions can be addressed by performing sensitivity forecasts for parameter inference from future cosmological data. The literature already presents many such forecasts (see e.g. \cite{Carbone:2010ik,Oyama:2012tq,Hamann:2012fe,Audren:2012vy,Pearson:2013iha,Allison:2015qca,Villaescusa-Navarro:2015cca,Oyama:2015gma,LoVerde:2016ahu,DiValentino:2016foa,Archidiacono:2016lnv,Schmittfull:2017ffw,Boyle:2017lzt,Sprenger:2018tdb,Hazra:2018eib,Yu:2018tem,Boyle:2018rva} and references therein). Previous works are usually based on different methods (e.g. Fisher matrix approaches with different prescriptions or Markov Chain Monte Carlo methods), different assumptions related to future observations (list of observables included for a given experiment, assumed instrumental sensitivities, list of systematic errors taken into account and marginalised over), and different cosmologies. 
To give examples, when doing forecasts for a given CMB experiment, one may choose whether to include information from CMB lensing extraction; for a given redshift survey, one may use different schemes to remove information from non-linear scales, to model bias and redshift space distortions, etc. 

For the sake of comparing the ability of different experiments to resolve parameter degeneracies, what is important in the present paper is not that our assumptions are the best (although we made an effort to implement as realistic assumptions as possible in our pipeline), but that they are the same across the variety of datasets and cosmological models that we consider. 
Thus, we present a three-dimensional grid of forecasts.
The three axes of the grid are: underlying cosmological models, with more or fewer free parameters; CMB experiments (or combination of them when relevant); Large Scale Structure surveys or combination of them.
We perform our forecasts with an MCMC exploration of the parameter space with a mock likelihood describing future data and instrumental sensitivities. For this we use the new version \texttt{3.0} of the cosmological inference package \texttt{MontePython}\footnote{https://github.com/brinckmann/montepython\_public}~\cite{Audren:2012wb,Brinckmann:2018cvx}, interfaced with the Boltzmann solver \texttt{CLASS}\footnote{http://class-code.net}~\cite{Blas:2011rf} \texttt{v2.7}. Using modern tools this method is reasonably fast and considerably more robust than Fisher matrix forecasts\footnote{
MCMC forecasts do not depend on an arbitrary choice of step size in the calculation of numerical derivatives, and avoid the numerical instabilities that often appear in numerical derivative calculations and in the inversion of high-dimensional matrices. Besides, they normally remain reliable when the likelihood deviates significantly from a multivariate Gaussian in the model parameters, and/or when there are strong degeneracies between parameters. In the worst case, in presence of a very strong degeneracy, MCMC forecast runs might have difficulties to converge. Then, one knows that there is an issue and may look for a solution (e.g. changing the parameter basis or fixing some parameters), which is better than quoting a possibly meaningless Fisher-based result.}. Nonetheless, executing a three-dimensional grid of MCMC parameter inference forecasts is a significant computational effort, hence we must carefully choose the number of cases that we want to consider for each axis of the grid.

As far as underlying models are concerned, we justify our choices in section \ref{models}. In summary, we will study the robustness of neutrino mass bounds against a variation of the parameters $N_\mathrm{eff}$, $w_0$, and $w_a$, but we will still assume that the universe obeys to Einstein's Theory of General Relativity and that it has negligible spatial curvature. 
On the side of CMB observations, we consider the current Planck sensitivity as a baseline, and discuss the improvement coming from four mission or survey projects: LiteBIRD, CORE-M5, CMB-S4 and PICO, for which we provide details in section \ref{experiments}. Finally, for LSS surveys, we restrict ourselves to the future BAO, galaxy redshift survey, cosmic shear survey, and intensity mapping survey that will be performed by DESI, Euclid, and SKA, also presented in more details in section \ref{experiments}. Finally, we discuss the impact of a precise measurement of the reionisation optical depth coming from future 21cm observations. 

The results of our forecasts are presented both in the form of compact figures in section \ref{results} and of extensive tables in appendix~\ref{appendix}. We only include in section \ref{results} the figures that describe the predicted sensitivity to $M_\nu$, but in principle all other forecasted parameter uncertainties can be interesting. Therefore, we produced similar figures for all other parameters; they can be downloaded from the public repository\newline
\url{https://brinckmann.github.io/montepython_public/neutrino_mass_forecasts/}.

%% file: models.tex
\section{Choice of models and parameter degeneracies\label{models}}

Our baseline model will be the minimal $\Lambda$CDM parametrised by 
$\{\omega_\mathrm{b}, \omega_\mathrm{cdm}, \theta_s, A_s, n_s, z_\mathrm{reio} \}$
(baryon and cold dark matter densities, angular scale of sound horizon, primordial amplitude and spectral index, redshift of reionization), extended to account for massive neutrinos. Cosmological observables are slightly sensitive to individual neutrino masses. For a fully realistic forecast, we should float the absolute neutrino mass scale and consider three different masses related to each other by the solar and atmospheric mass square differences, according to the normal or inverted hierarchy~\cite{Gariazzo:2018pei}. However, it has been shown that the difference between cosmological observables for these realistic neutrino mass hierarchies and a mass-degenerate model with the same total mass $M_\nu$ is extremely small (at most 0.1\% in the matter power spectrum) and below the sensitivity of future experiments \cite{Lesgourgues:2004ps,Lesgourgues:2006nd,Lesgourgues:1519137}\footnote{These references also show that such conclusions would not extend to models with one massive and two massless or two massive and one massless species.}. As such, for simplicity we stick to the mass-degenerate model and consider three massive species with masses $M_\nu/3$ each.

Even in the context of a simple $\Lambda$CDM cosmology, the neutrino mass has non-trivial correlations with some of the six free model parameters. These correlations have been discussed recently in \cite{Archidiacono:2016lnv} for various combinations of future datasets. It appears that a significant correlation between $M_\nu$ and the optical depth (or redshift) of reionization, although absent in fits of CMB data only, will appear when combining future CMB, BAO, and LSS data sets. Thus, better measurements of $\tau_\mathrm{reio}$ using additional techniques will allow to tighten neutrino mass bounds. This is the reason for which we will consider later a hypothetical measurement of $\tau_\mathrm{reio}$ based on 21cm intensity mapping as a complement to CMB, BAO, and LSS data sets.

Next, we want to concentrate on extended cosmologies known to bring potential parameter degeneracies with $M_\nu$ and to increase the standard deviation $\sigma(M_\nu)$. The literature often discusses potential degeneracies of $M_\nu$ with extra relativistic relics parametrised by $N_\mathrm{eff}$, with dynamical dark energy with an equation of state parameter $p_\mathrm{DE} (z)/\rho_\mathrm{DE} (z) = w(z)$, with spatial curvature parametrised with $\Omega_k$, and with models of modified gravity (beyond Einstein's theory of General Relativity). However, we cannot realistically cover all cases, but let us discuss them one by one. 
\begin{itemize}
\item
The $M_\nu$--$N_\mathrm{eff}$ parameter degeneracy was a known concern with old datasets \cite{Hannestad:2003xv,Elgaroy:2003yh,Hannestad:2003ye,Crotty:2004gm,Archidiacono:2013lva}, but all post-Planck fits have shown that it is now resolved. Nonetheless, we will study eight-parameter models with ($M_\nu$,\,$N_\mathrm{eff}$), in order to show once again that the neutrino mass measurement will be stable against varying $N_\mathrm{eff}$. Furthermore, $N_\mathrm{eff}$ is {\it per se} a particularly interesting parameter to fit to the data, given the wide range of models with new particle physics assumptions that it covers.
\item
Dynamical dark energy should, in principle, be described by an infinity of parameters: even in the sub-class of models in which DE perturbations are negligible, the DE background evolution can be described by a free function $w(z)$. However, unless one is interested in early dark energy models, this function impacts cosmological observables only at small redshift, and a CPL parametrisation \cite{Chevallier:2000qy} $w(z)= w_0 + w_a (1-a/a_0)$ with two free parameters $w_0$ and $w_a$ is usually sufficient to catch the main features of a given model and study degeneracies with other parameters. We will stick to this case, and we include phantom values ($w(z)<-1$) already known to result in weaker bounds on $M_\nu$ (see e.g. \cite{DiValentino:2016foa}).
\item
Spatial curvature is partially degenerate with neutrino masses because both $M_\nu$ and $\Omega_k$ affect the CMB peak scale; this degeneracy, however, is lifted by other neutrino mass effects, such as neutrino free-streaming affecting the matter growth factor after the non-relativistic transition. This case is definitely interesting, but in order to keep the computational effort tractable, we defer it to a later study.
\item
Modifications of gravity theories beyond Einstein's General Relativity are a Pandora's box, since phenomenological parametrisations contain a large number of free functions. Some particular theories could introduce a scale-dependent growth factor of non-relativistic matter fluctuations that would counteract the massive neutrino effects at each redshift. Several authors have studied the degeneracy between $M_\nu$ and some extended gravity parameters for several particular cases, and found that parameter degeneracies are present or not depending on the assumed gravity model and mock data set
(see e.g. \cite{Motohashi:2010sj,Motohashi:2012wc,Baldi:2013iza,He:2013qha,Barreira:2014ija,Hu:2014sea,Shim:2014uta,Bellomo:2016xhl,Alonso:2016suf}). We will not explore this direction in the present paper.
\end{itemize}
In summary, we will study the robustness of neutrino mass bounds against floating the parameters $N_\mathrm{eff}$, $w_0$, and $w_a$, but we will still assume that the universe obeys to Einstein's theory of General Relativity and that it has negligible spatial curvature. As usual in minimal cosmology scenarios, we perform several assumptions on the particle content of the universe: dark matter is cold and collisionless, neutrinos are non-interacting and have a frozen thermal distribution, etc. These assumptions are motivated by simplicity, but relaxing them could also be a way to relax neutrino mass bounds. Finally, we assume a power-law spectrum of primordial curvature perturbations and a negligible spectrum of tensor perturbations, but these assumptions are harmless for neutrino mass measurements: a modification of initial conditions could not be degenerate with neutrino mass effects, given that it affects all observables in the same way at all times, while the effect of neutrino masses is notoriously dynamical, depending on redshift and on the types of observables \cite{Lesgourgues:2006nd,Lesgourgues:1519137}.

The mock likelihoods discussed in the next section assume three degenerate massive neutrinos, no extra relativistic degrees of freedom and a fiducial model consistent with Planck\footnote{The fiducial model is given by the following parameter values:\\
$\omega_\mathrm{b}=0.02218$, $\omega_\mathrm{cdm}=0.1205$, $\theta_s=1.04146$, $\ln10^{10}A_s=3.056$, $n_s=0.9619$, $z_\mathrm{reio}=8.24$, taken from table 8, column 5 of \cite{Aghanim:2016yuo}. }.
 

%% file: experiments.tex
\section{Experimental sensitivities and mock likelihoods \label{experiments}}

We employ the MCMC forecast method detailed in \cite{Perotto:2006rj}: a future experiment is encoded as a mock likelihood, providing the probability that the mock data is true given the model assumed at each step of the MCMC parameter exploration. We neglect scattering in the mock data: it is directly given by the $C_\ell$s of the assumed fiducial model. Generating $a_{\ell m}$s as a single random realisation of the fiducial theory would change the reconstructed means, but not the sensitivities \cite{Perotto:2006rj} (except in very rare and unlikely realisations). In the case of mock likelihoods accounting for large scale structure, we marginalise over nuisance parameters that account for residual systematic effects. However, the results of such forecasts should always be taken with care: in reality, the modelling of systematics is often slightly incorrect or incomplete.

For CMB experiments, we assume a Gaussian likelihood for the multipole coefficients of temperature, polarisation, and CMB lensing potential maps, described by equations (3.1) to (3.7) of \cite{Perotto:2006rj}. The noise spectra of temperature and polarisation are inferred from the resolution and sensitivity parameters that are expected to reflect the characteristics of the instruments according to standard approximations (see e.g. equation (2.2) in \cite{Perotto:2006rj}). The temperature and polarisation noises are assumed to be statistically independent, which means that the noise spectrum $N_\ell^{TE}$ is approximated as zero. Given the fiducial model and noise spectra, one can estimate the error that would be performed on the measurement of the lensing potential spectrum by running a quadratic estimator \cite{Okamoto:2003zw}. These estimators are based on products of four multipoles, each of the T, E, or B type, but for a more conservative forecast we discard any information coming from the auto-correlation of the B-mode maps, due to the non-Gaussianity of the $a_{\ell m}^B$ multipoles. All quadratic estimators can then be combined in order to minimise the reconstruction noise: this defines the minimum variance estimator \cite{Okamoto:2003zw}. In this paper we use the FuturCMB\footnote{\url{http://lpsc.in2p3.fr/perotto/}} code \cite{Perotto:2006rj} to compute the noise spectrum of the CMB lensing potential reconstructed with the minimum variance estimator.

We are interested in the comparison of four future CMB experiments, based on current expectations for their instrumental sensitivity.
These are: 
\begin{itemize}
\item
The LiteBIRD\footnote{\url{http://litebird.jp/eng/}} satellite project of JAXA~\cite{Matsumura:2013aja,Suzuki:2018cuy}, currently in phase A and optimised for primordial B-modes, with very good sensitivity but modest resolution.
\item
The CORE-M5 satellite project~\cite{Delabrouille:2017rct}, recently submitted to the M5 call of ESA and not approved within this call, but still being considered for future applications: CORE-M5 would have a slightly better sensitivity and significantly better resolution than LiteBIRD.
\item
The CMB Stage Four\footnote{\url{https://cmb-s4.org}} (CMB-S4) project~\cite{Abazajian:2016yjj,Abitbol:2017nao}, an ambitious project gathering many ground-based detectors that should be deployed over the next decade, with outstanding resolution and sensitivity but smaller sky coverage than satellites.
\item
The PICO satellite project\footnote{See 
\url{https://zzz.physics.umn.edu/ipsig/start} and
\url{https://zzz.physics.umn.edu/ipsig/_media/pico_science_aas_v11.pdf}; 
channel resolution and sensitivity taken from \url{https://zzz.physics.umn.edu/ipsig/baseline}.}~\cite{Sutin:2018onu,Young:2018aby}
that may be submitted to NASA in the future, and which would improve over the sensitivity of LiteBIRD by a factor of 3 to 4.
\end{itemize}
For each experiment and each channel we assume a resolution, a sensitivity, and a sky fraction summarised in Table~\ref{tab:ext_cmb_specs}.
Given the high degree of complementarity between satellite and ground-based missions, as they are optimised respectively for large and small angular scales, it is natural to combine them. We will study two possible combinations: LiteBIRD + CMB-S4 and CORE-M5 + CMB-S4. For simplicity, we do not consider regions of overlapping data. We thus make the same assumption as in \cite{Abazajian:2016yjj}: we consider that the optimal combination will consist in LiteBIRD/CORE-M5 data for $\ell \leq 50$, CMB-S4 data for $\ell > 50$ in the region covered by the experiment (40\% of the sky), and additional high-$\ell$ data from LiteBIRD/CORE-M5 in the region covered by the satellite but not by CMB-S4 (30\% of the sky).
\begin{table} [h!]
	\begin{center}\footnotesize
		\begin{tabular}{|c|c|c|c|}
			\hline
			Channel [GHz] & FWHM [arcmin] & $\Delta T $ [$\mu$K arcmin] & $\Delta P $ [$\mu$K arcmin] \\ 
			\hline
			\hline
			\multicolumn{4}{|c|}{1. LiteBIRD,  $\ell_{\rr{max}} = 1350, f_{\rr{sky}} = 0.7$  }    \\
			\hline
			$140$ & $31$  & $4.1 $ & $5.8$ \\
			\hline
			\hline
			\multicolumn{4}{|c|}{2. CORE-M5,  $\ell_{\rr{max}} = 3000, f_{\rr{sky}} = 0.7$  }    \\
			\hline
			$130$ & $8.51$  & $3.9$ & $5.5$ \\
			$145$ & $7.68$  & $3.6$ & $5.1$ \\
			$160$ & $7.01$  & $3.7$ & $5.2$ \\
			$175$ & $6.45$  & $3.6$ & $5.1$ \\
			$195$ & $5.84$  & $3.5$ & $4.9$ \\
			$220$ & $5.23$  & $3.8$ & $5.4$ \\
			\hline
			\hline
			\multicolumn{4}{|c|}{3. CMB-S4,  $\ell_{\rr{min}} = 30, \ell_{\rr{max}} = 3000, f_{\rr{sky}} = 0.4$  }    \\
			\hline
			$150$ & $3.0$  & $1.0 $ & $1.41$ \\
			\hline			
			\hline
			\multicolumn{4}{|c|}{4. LiteBIRD + CMB-S4 in combination} \\
			\hline
			\multicolumn{4}{|c|}{low-$\ell$ from LiteBIRD,  $\ell_{\rr{max}} = 50, f_{\rr{sky}} = 0.7$  }    \\
			\hline
			$140$ & $31$  & $4.1 $ & $5.8$ \\
	                 \hline
			\multicolumn{4}{|c|}{high-$\ell$ from CMB-S4,  $\ell_{\rr{min}} = 51, \ell_{\rr{max}} = 3000, f_{\rr{sky}} = 0.4$  }    \\
			\hline
			$150$ & $3.0$  & $1.0 $ & $1.41$ \\
			\hline
			\multicolumn{4}{|c|}{additional high-$\ell$ from LiteBIRD,  $\ell_{\rr{min}} = 51, \ell_{\rr{max}} = 1350, f_{\rr{sky}} = 0.3$  }    \\
			\hline
			$140$ & $31$  & $4.1 $ & $5.8$ \\
			\hline
			\hline
			\multicolumn{4}{|c|}{ 5. CORE-M5 + CMB-S4 in combination} \\
			\hline
			\multicolumn{4}{|c|}{low-$\ell$ from CORE-M5,  $\ell_{\rr{max}} = 50, f_{\rr{sky}} = 0.7$  }    \\
			\hline
			\hline
			$130$ & $8.51$  & $3.9$ & $5.5$ \\
			$145$ & $7.68$  & $3.6$ & $5.1$ \\
			$160$ & $7.01$  & $3.7$ & $5.2$ \\
			$175$ & $6.45$  & $3.6$ & $5.1$ \\
			$195$ & $5.84$  & $3.5$ & $4.9$ \\
			$220$ & $5.23$  & $3.8$ & $5.4$ \\
	        \hline
			\multicolumn{4}{|c|}{high-$\ell$ from CMB-S4,  $\ell_{\rr{min}} = 51, \ell_{\rr{max}} = 3000, f_{\rr{sky}} = 0.4$  }    \\
			\hline
			$150$ & $3.0$  & $1.0 $ & $1.41$ \\
			\hline
			\multicolumn{4}{|c|}{additional high-$\ell$ from CORE-M5,  $\ell_{\rr{min}} = 51, \ell_{\rr{max}} = 3000, f_{\rr{sky}} = 0.3$  }    \\
			\hline
			$130$ & $8.51$  & $3.9$ & $5.5$ \\
			$145$ & $7.68$  & $3.6$ & $5.1$ \\
			$160$ & $7.01$  & $3.7$ & $5.2$ \\
			$175$ & $6.45$  & $3.6$ & $5.1$ \\
			$195$ & $5.84$  & $3.5$ & $4.9$ \\
			$220$ & $5.23$  & $3.8$ & $5.4$ \\
			\hline
			\hline
			\multicolumn{4}{|c|}{6. PICO, $\ell_{\rr{max}} = 3000, f_{\rr{sky}} = 0.7$  }   \\
			\hline
			$62.2$ & $12.8$  & $2.76$ & $3.9$ \\
			$74.6$ & $10.7$  & $2.26$ & $3.2$ \\
			$89.6$ & $9.5$  & $1.41$ & $2.0$ \\
			$107.5$ & $7.9$  & $1.20$ & $1.7$ \\
			$129.0$ & $7.4$  & $1.13$ & $1.6$ \\
			$154.8$ & $6.2$  & $0.99$ & $1.4$ \\
			$185.8$ & $4.3$  & $1.84$ & $2.6$ \\
			$222.9$ & $3.6$  & $2.19$ & $3.1$ \\
			\hline
		\end{tabular}
	\end{center}
	\caption{Experimental specifications for the CMB experiments used in this work. From left to right, frequency channel(s) dedicated to cosmology, beam width, temperature and polarisation sensitivities for this/these channel(s). See the text for references to each experiment or combination of experiments.}
	\label{tab:ext_cmb_specs}
\end{table}

It is useful to add Planck to this list of experiments, in order to quantify the progress that can be made compared to the present situation. We choose not to use the actual Planck likelihood, in order to benefit from a feature of \texttt{MontePython}: when running only mock likelihoods, the code automatically creates mock spectra at the beginning of the first run (for fiducial parameter values specified by the user), with, of course, a single fiducial model being used across all likelihoods. Therefore, in our grid of forecast, it is technically easier to use a mock Planck likelihood, with resolution and sensitivity assumptions close to those of the full Planck mission~\cite{Aghanim:2018eyx}. It is impossible to mimic exactly the real sensitivity of the Planck results, due to our assumption of a Gaussian CMB likelihood with uncorrelated $a_{\ell m}$s (breaking mainly at low $\ell$s), of uncorrelated temperature and polarisation noise, and of perfect foreground cleaning up to $\ell_\mathrm{max}$. Nevertheless, we made an educated guess for the noise level in our mock Planck likelihood leading to sensitivities very close to the real ones\footnote{With the exception of the error $\sigma(\tau_\mathrm{reio})$ on the optical depth, which is still smaller in our forecast than in reality - but this does not propagate to the error $\sigma(M_\nu)$ on the summed neutrino mass, since in our ``Planck alone'' runs, $\sigma(M_\nu)$ is still within 8\% of the true Planck 2018 results \cite{Aghanim:2018eyx}.}.

\begin{figure}
	\centering
	\includegraphics[width=5.2cm]{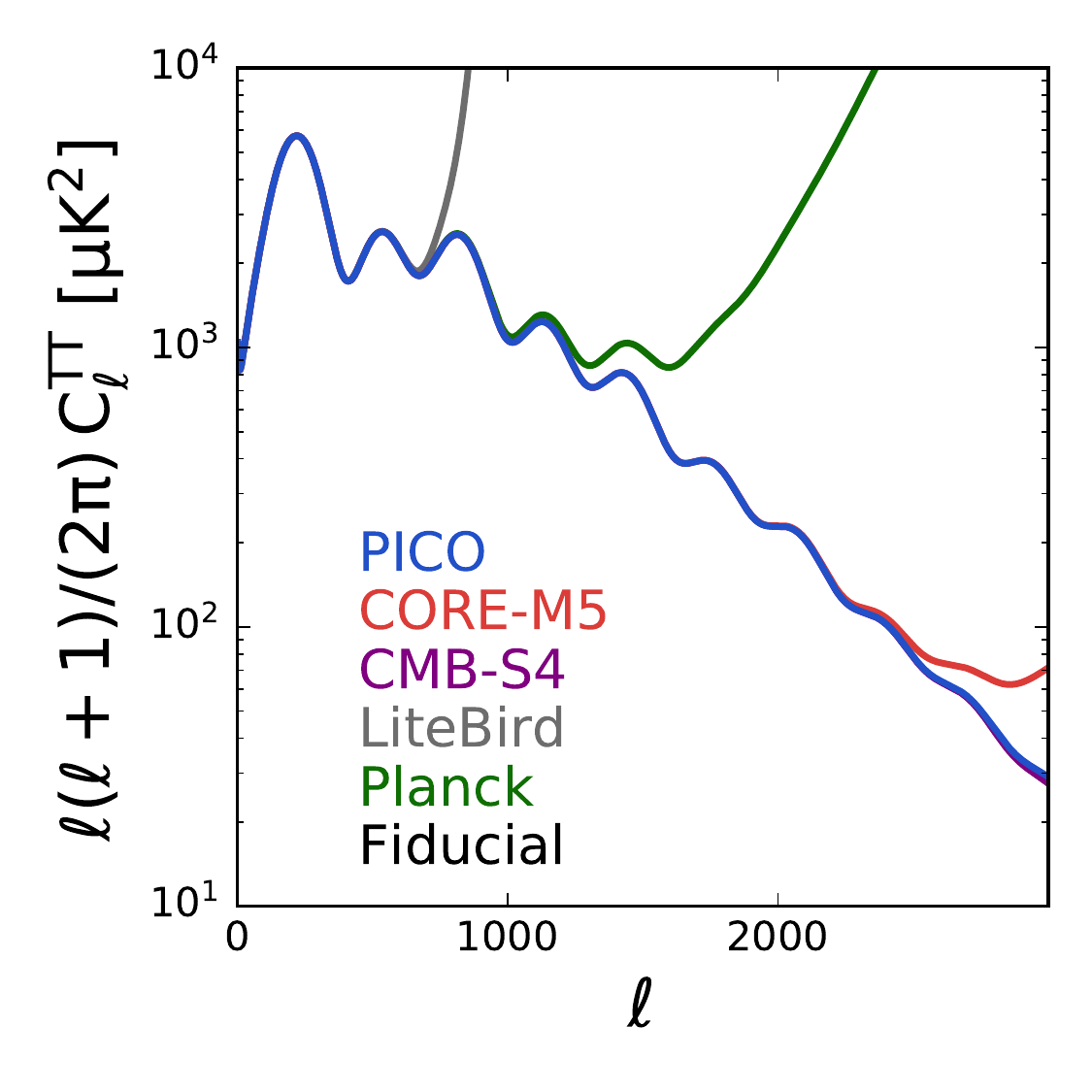} 
	\hspace{-0.2cm}\includegraphics[width=5.2cm]{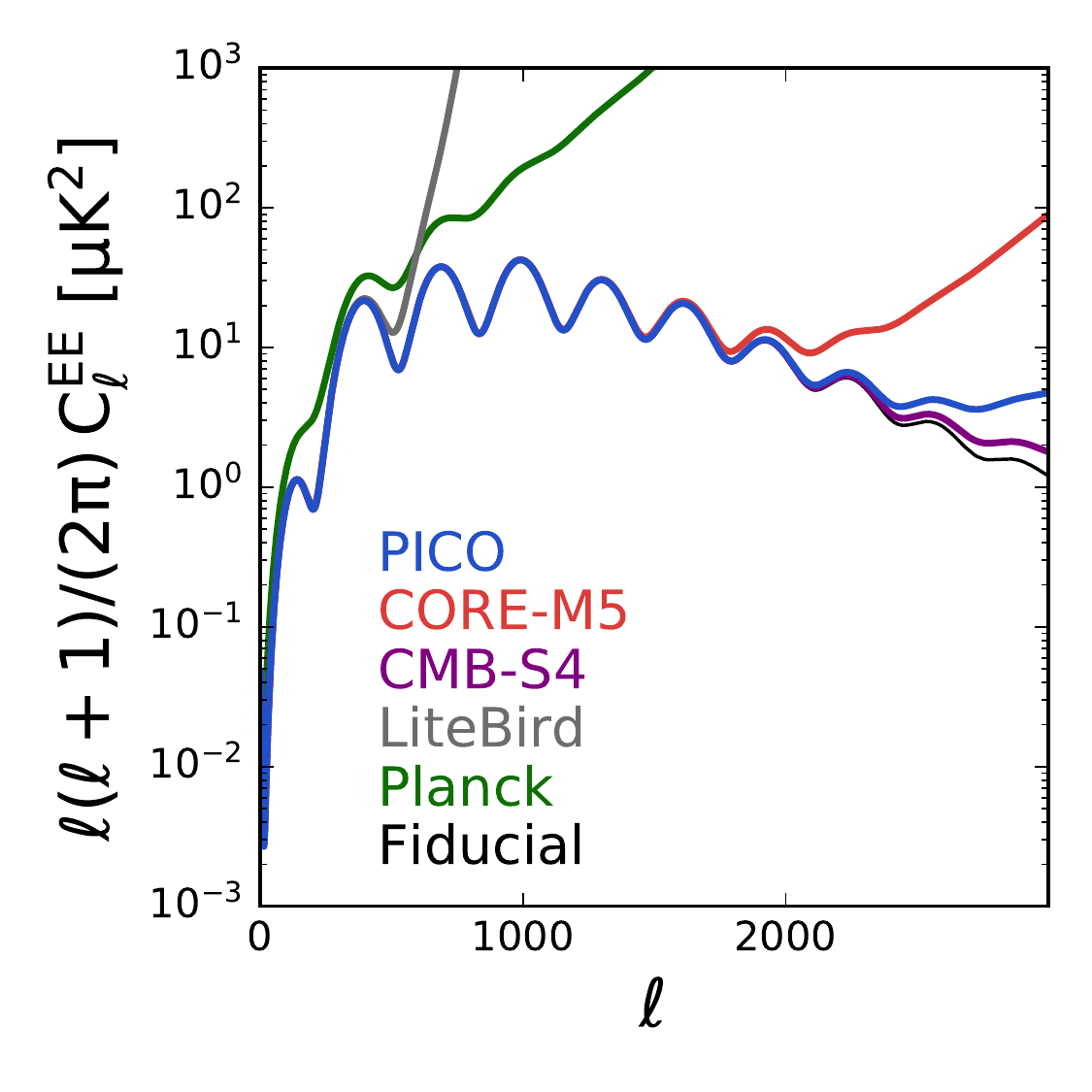}
	\hspace{-0.2cm}\includegraphics[width=5.2cm]{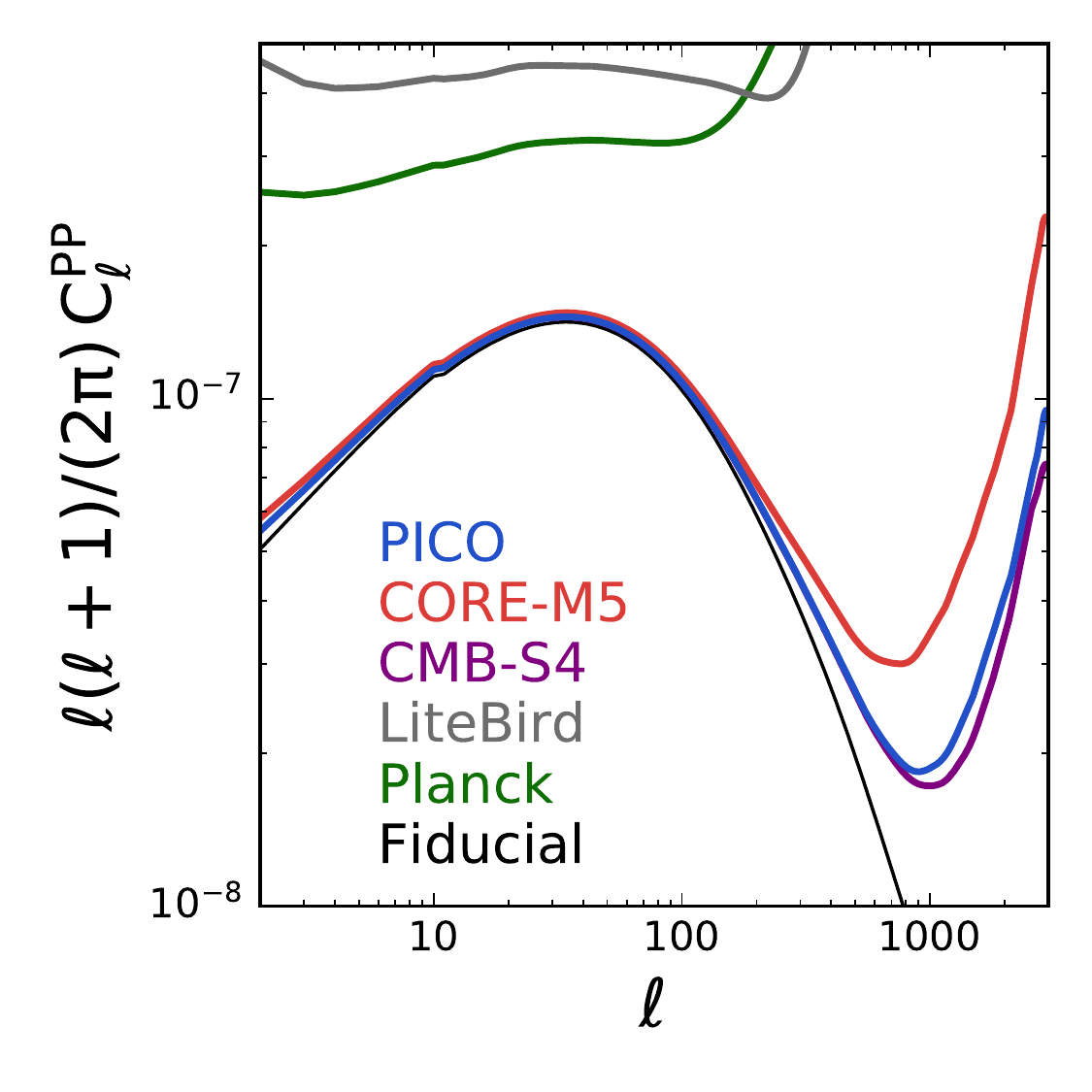}	
	\caption{The figure shows the fiducial model (black) compared to the model plus noise spectrum based on the minimum variance estimator, $C_l + N_l$, for each experiment (Planck in green, LiteBIRD in grey, CMB-S4 in purple, CORE-M5 in red, and PICO in blue), for temperature anisotropies (left), E-mode polarisation (middle), and CMB lensing potential (right). CMB-S4 compares favourably to PICO, but it is important to note that the noise depicted here does not include cosmic variance or uncertainties due to imperfect foreground removal, where PICO would have a clear advantage over CMB-S4 due to full-sky observations and a much larger number of channels.}
	\label{fig:noise}
\end{figure}

The noise spectra computed for these experiments using the minimum variance estimator are shown in \cref{fig:noise}. As designed, the role of LiteBIRD is mainly on large scales for polarisation, where the experiment does very well. However, LiteBIRD does not add significantly to CMB lensing information. CMB-S4 does extraordinarily well (except, obviously, on very large scales), but we remind the reader that this information neglects the small sky fraction (i.e. this noise is computed only based on the sensitivity and beam size), and foreground issues, which might be more pronounced for a ground based experiment with less sky coverage and fewer channels available than a satellite mission. The CORE-M5 sensitivity is essentially the same as that presented in the CORE parameters paper~\cite{DiValentino:2016foa}, with a slight improvement for CMB lensing extraction coming from the use of the minimum variance estimator noise spectra instead of the EB estimator. In the end, once sky coverage is taken into account, the futuristic PICO satellite should out-perform CMB-S4, which covers a much smaller fraction of the sky, while also having the advantage of a large number of channels and full-sky observations for improved foreground removal.

\vspace{1cm}
 
\noindent For LSS surveys, we will focus on three upcoming missions and data sets:
\begin{itemize}
\item BAO data from the Dark Energy Spectroscopic Instrument (DESI), starting observations in 2019. DESI is designed to measure the BAO scale in the redshift range $0.05 <z< 2.1$, but we restrict ourselves to $0.15 < z < 1.85$ to be conservative, with percent-level precision~\cite{Levi:2013gra,Aghamousa:2016zmz}.
\item Galaxy clustering and cosmic shear data from the Euclid satellite~\cite{Amendola:2012ys,Amendola:2016saw}, scheduled to be launched in 2021. Euclid is set to perform an incredibly precise galaxy survey out to $z>2$, aiming at a 1\% accuracy on the galaxy clustering and cosmic shear observables.
\item Intensity mapping from the Square Kilometre Array (SKA)~\cite{Santos:2015gra}. SKA will be the world's largest radio telescope, and will not only provide cosmic shear and galaxy clustering information, but also produce a map of neutral hydrogen through 21cm intensity mapping, allowing us to trace the LSS distribution up to redshift $z\sim 20$.
\end{itemize}

\noindent For DESI BAO we use the same configuration as in~\cite{Archidiacono:2016lnv}.
For Euclid and SKA, we rely on exactly the same approach and mock likelihoods as in~\cite{Sprenger:2018tdb}\footnote{The likelihoods from~\cite{Sprenger:2018tdb} were made public with \texttt{MontePython v3.1.}}. We summarise the most relevant assumptions below, and refer the reader to~\cite{Sprenger:2018tdb} for further details.
\begin{itemize}
\item{\bf Galaxy clustering from Euclid and intensity mapping from SKA.} 
For those two observables, we quantify the information in terms of three dimensional power spectrum, including redshift-space distortions~\cite{Kaiser:1987qv}, fingers of God~\cite{Bull:2014rha} and Alcock-Paczynski effects.
We consider galaxies and hydrogen as biased tracers of the cold dark matter + baryon field only, rather than of the total matter field (i.e., including the contribution of massive neutrinos). Given this consideration, we correct for the scale-dependent neutrino induced bias~\cite{Villaescusa-Navarro:2013pva,Castorina:2013wga,Costanzi:2013bha,Biagetti:2014pha,Castorina:2015bma}. This effect was recently shown to be crucial when measuring massive neutrinos with future surveys like Euclid~\cite{Raccanelli:2017kht,Vagnozzi:2018pwo}. Note, however, that even with this correction a residual scale-dependent bias remains~\cite{LoVerde:2014pxa,LoVerde:2014rxa,Chiang:2017vuk,Munoz:2018ajr,Chiang:2018laa}.
Concerning non-linear corrections, we use \texttt{HALOFIT}~\cite{Takahashi:2012em}. We neglect the neutrino corrections of~\cite{Bird:2011rb}, because, as we have just explained, the source is the cold dark matter + baryon field only. We also take into account residual uncertainties in the bias modeling by marginalising over two nuisance parameters. Moreover, we implement a theoretical error that takes into account our imperfect knowledge of non-linear modeling, including further residual errors on the bias, and effectively replaces the non-linear cut-off commonly used in the literature. The envelope of the error increases gradually with wavenumber, from a fixed 0.33\% below $k=0.01 \ h/Mpc$, to 1\% at $k=0.3 \ h/Mpc$, and 10\% at $k=10 \ h/Mpc$, at which point a cut-off is applied (this configuration is labeled as \textit{realistic} in~\cite{Sprenger:2018tdb}).
For intensity mapping, we focus on the SKA1 band 2 survey, executed in single dish mode, with 10000 hours of observations and 200 dishes, probing the range $0.05 \leq z \leq 0.45$ (divided in 4 redshift bins) across 58\% of the sky. For Euclid we use 13 redshift bins probing $0.7 \leq z \leq 2.0$, so there is no overlap with SKA.
\item {\bf Cosmic shear from Euclid.}
For cosmic shear, we construct an angular harmonic power spectrum as our mock dataset. We assume Limber and flat-sky approximations~\cite{Lemos:2017arq,Asgari:2016txw}.
Since weak gravitational lensing is sourced by the total matter field, we use here the \texttt{HALOFIT}~\cite{Takahashi:2012em} non-linear corrections including the neutrino contribution of~\cite{Bird:2011rb}. 
As explained in \cite{Sprenger:2018tdb}, we account for non-linear uncertainties by introducing a bin-dependent cut-off in multipole space, $\ell_{\mathrm{max}}^i = k_{\mathrm{NL}}(\bar{z}^i) \times \bar{r}_{\mathrm{peak}}^i$, inferred from the redshift-dependent scale of non-linearity $k_{\mathrm{NL}}(z) = k_{\mathrm{NL}}({z=0}) \times (1+z)^{2/(2+n_s)}$,
and from the comoving distance to the peak of the window function $\bar{r}_{\mathrm{peak}}^i$ in bin $i$
(this configuration is labeled as \textit{conservative} in~\cite{Sprenger:2018tdb}).
\end{itemize}

\noindent As shown in~\cite{Liu:2015txa,Villaescusa-Navarro:2015cca} and explained in~\cite{Archidiacono:2016lnv}, the combination of future CMB, BAO, and LSS data sets will lead to neutrino mass bounds correlated with other parameters like $H_0$, $A_s$, $n_s$, and $\tau_\mathrm{reio}$. Thus, any further improvements of the bounds on the latter parameters would result in a better determination of the neutrino mass. However, the best constraints on $H_0$, $A_s$, and $n_s$ may come from the same data combination, and will probably be difficult to improve. For $\tau_\mathrm{reio}$, there is some hope to gather additional independent information by exploiting the power of 21cm intensity mapping. For instance, HERA or SKA could achieve a measurement accurate up to $\sigma(\tau_\mathrm{reio})=0.001$~\cite{Liu:2015txa,Villaescusa-Navarro:2015cca}, if the astrophysical uncertainties are under control. We will add a Gaussian $\tau_\mathrm{reio}$--prior with such a standard deviation as a final mock likelihood in order to study the impact of only one of the information that these experiments can provide (since 21cm surveys will not only measure the evolution of the mean free electron fraction, but also the power spectrum of the 21cm signal at different redshifts \cite{Villaescusa-Navarro:2015cca}).

%% file: results.tex
\section{Results \label{results}}

\begin{figure}[htb]
	\centering
	\includegraphics[width=1.0\linewidth]{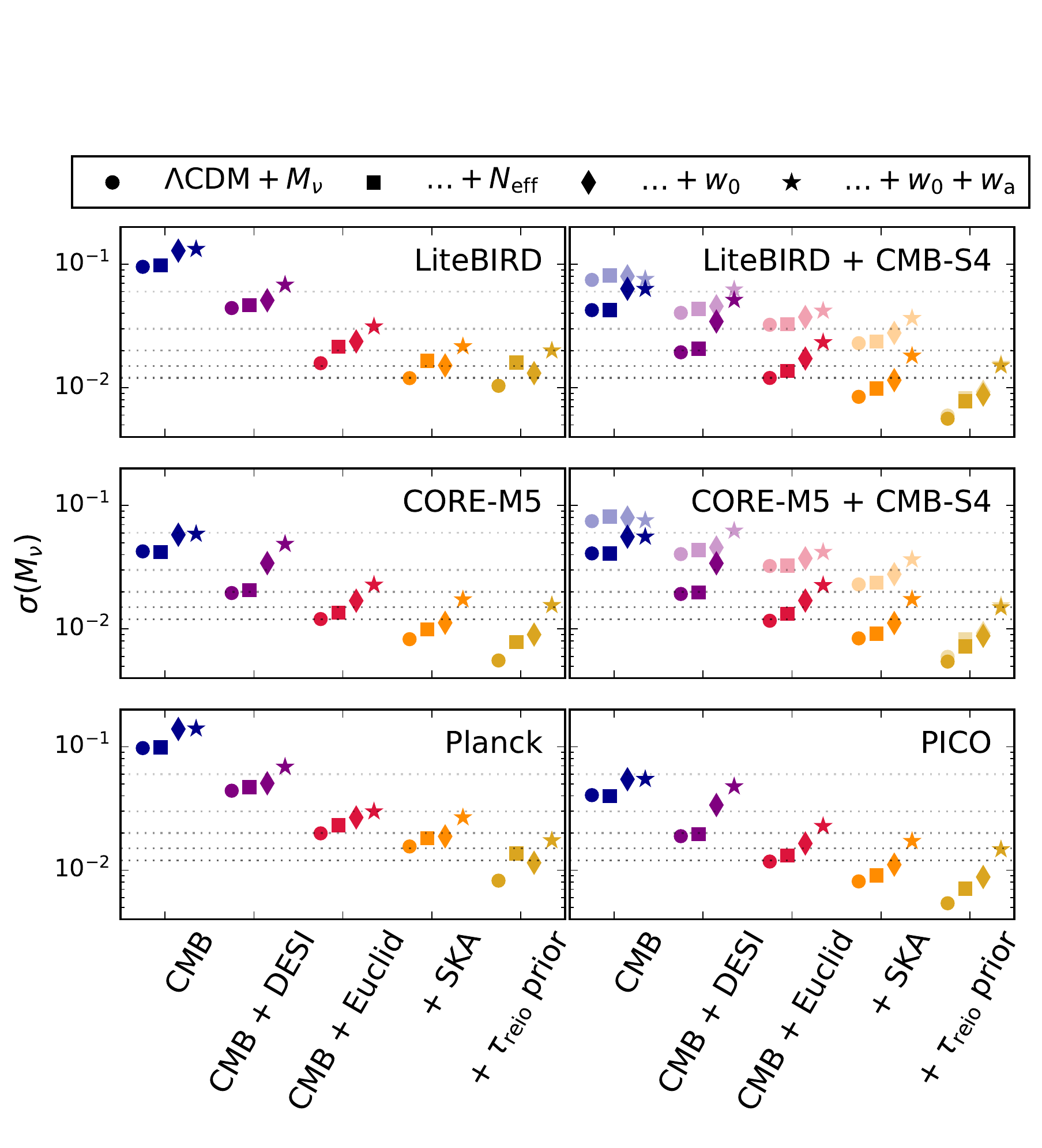}
	\caption{Neutrino mass sensitivity for each CMB experiment, alone and in combination with DESI, Euclid, Euclid + SKA1 IM, Euclid + SKA1 IM + $\tau_{\rm reio}$ prior. Each subplot corresponds to one CMB setup (LiteBIRD, LiteBIRD + CMB-S4, CORE-M5, CORE-M5 + CMB-S4, Planck, or PICO from top left to bottom right, where the desaturated symbols indicate the CMB-S4 sensitivity) and relevant combinations with large-scale structure surveys (reported on the x-axis). For each combination the sensitivity is depicted for four cosmological models: the minimal scenario $\Lambda$CDM $+M_\nu$, and three extensions $+N_{\rm eff}$, $+w_0$, and $+w_0+w_a$. The horizontal dashed lines show the thresholds for a 1 to 5$\sigma$ detection of $M_\nu=0.06$~eV.}
	\label{fig:Mnu_sensitivity}
\end{figure}

\begin{figure}[htb]
	\centering
	\includegraphics[width=1.0\linewidth]{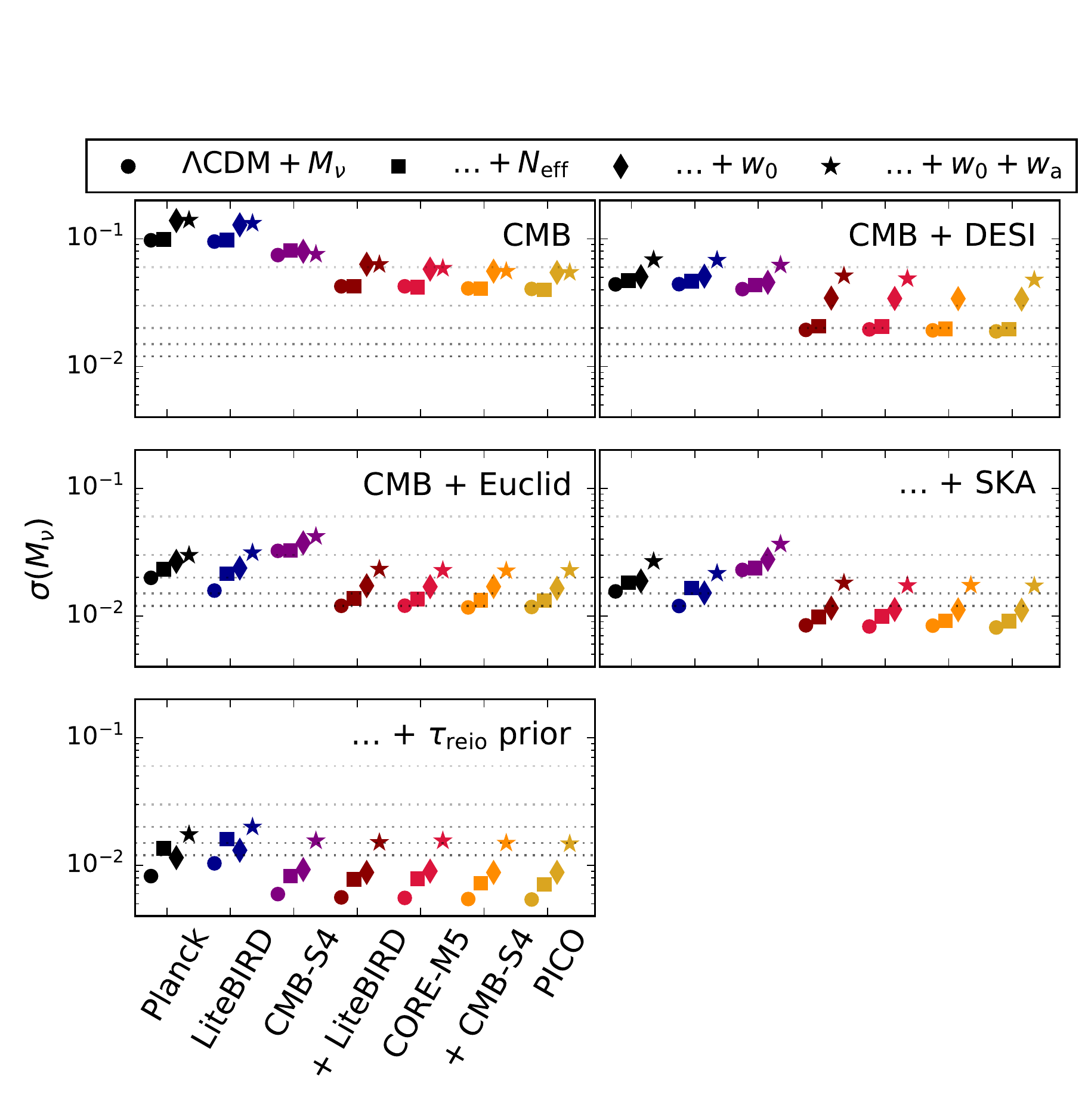}
	\caption{Neutrino mass sensitivity for each CMB experiment, alone and in combination with DESI, Euclid, Euclid + SKA1 IM, Euclid + SKA1 IM + $\tau_{\rm reio}$ prior. Each subplot corresponds to one CMB setup plus large-scale structure survey combination (CMB only, CMB + DESI, CMB + Euclid, CMB + Euclid + SKA1 IM, CMB + Euclid + SKA1 IM + $\tau_{\rm reio}$ prior from top left to bottom right). The horizontal dashed lines show the thresholds for a 1 to 5$\sigma$ detection of $M_\nu=0.06$~eV.}
	\label{fig:cmb_Mnu_sensitivity}
\end{figure}

The complete results of our forecasts are summarised in six tables in appendix \ref{appendix}, \cref{tab:planck,tab:litebird,tab:s4,tab:s4+litebird,tab:core,tab:s4+core,tab:pico}. A more intuitive and graphical summary of the sensitivity to
the neutrino mass $\sigma(M_\nu)$, which is the main focus of this work, is presented in \cref{fig:Mnu_sensitivity,fig:cmb_Mnu_sensitivity}.
We produced similar plots for all other parameters, which interested readers can download from the \texttt{MontePython} website\footnote{\url{https://brinckmann.github.io/montepython_public/neutrino_mass_forecasts/}}.

In \cref{fig:Mnu_sensitivity}
the results are ordered in a way to highlight the impact of each LSS dataset in combination with a given CMB experiment, whereas \cref{fig:cmb_Mnu_sensitivity}
shows the importance of using more precise CMB datasets in combination with a given LSS experiment.

\begin{figure}[htb]
	\centering
	\includegraphics[width=0.7\linewidth]{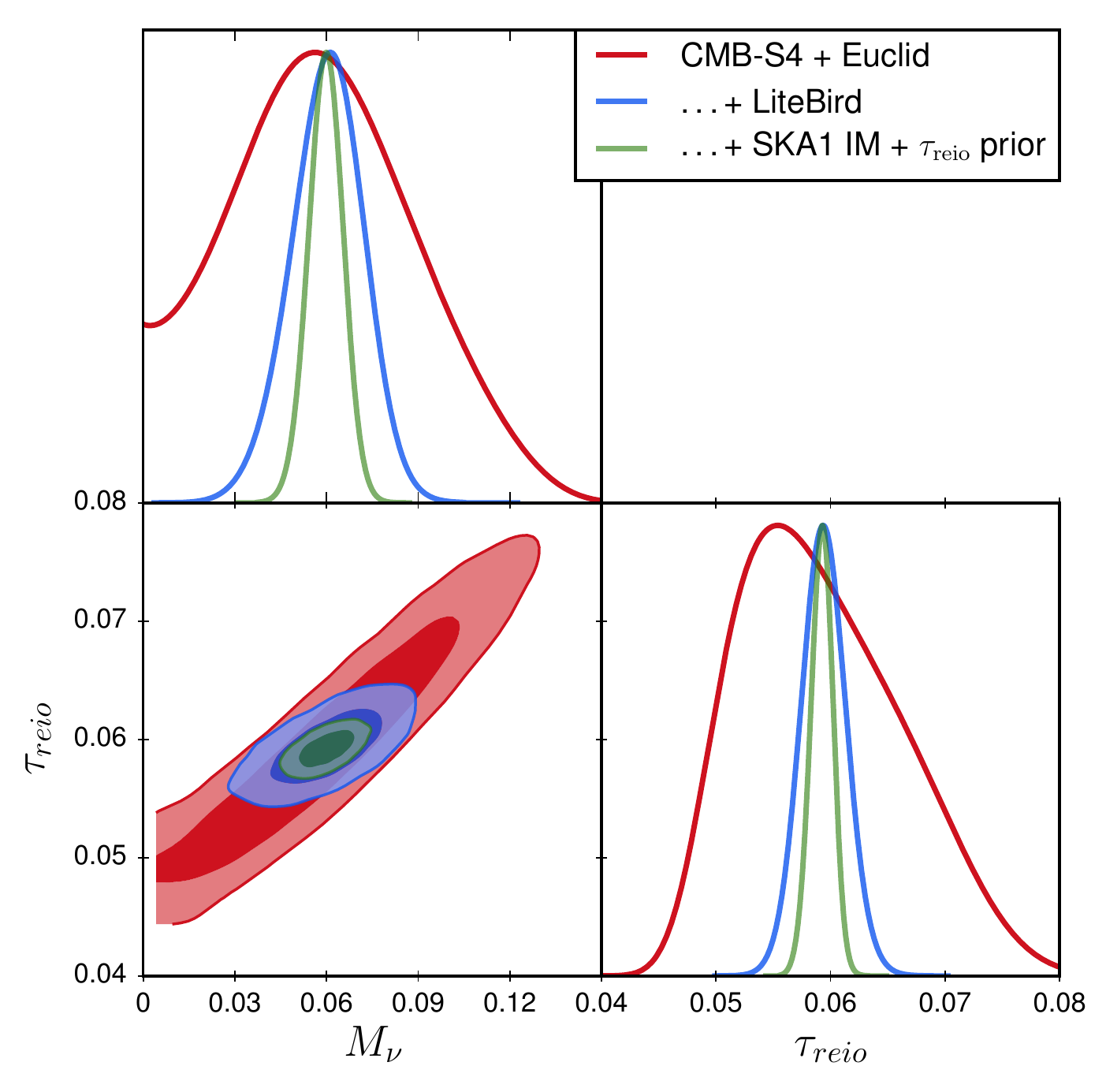}
	\caption{Neutrino mass vs $\tau_{\rm reio}$ for the three configurations CMB-S4 + Euclid, CMB-S4 + Euclid + LiteBIRD, CMB-S4 + Euclid + SKA1 IM + $\tau_{\rm reio}$ prior in the minimal 7 parameter $\Lambda$CDM+$M_{\nu}$ model.}
	\label{fig:cmb-s4_mnu-tau}
\end{figure}

First let us consider the minimal cosmological scenario ($\Lambda$CDM + $M_{\nu}$). \Cref{fig:Mnu_sensitivity,fig:cmb_Mnu_sensitivity} show that, in this case, for a fiducial neutrino mass sum $M_\nu=0.06$~eV:
\begin{itemize}
	\item CORE-M5 and PICO are so sensitive that they would only need to be combined with the BAO scale data from DESI for a 3-$\sigma$ detection,
	\item a 3- to 4-$\sigma$ detection could be achieved already by Planck or LiteBIRD when combined with Euclid,
	\item LiteBIRD in combination with Euclid and SKA1 intensity mapping reaches the 5-$\sigma$ threshold, 
	\item CORE-M5 or PICO would also achieve a 5-$\sigma$ detection in combination with Euclid only, 
	\item CORE-M5 or PICO would even achieve a 7-$\sigma$ detection when SKA1 intensity mapping data is added, and a 10-$\sigma$ one if $\tau_{\rm reio}$ could be further constrained according to our assumed prior. This illustrates the enormous benefit towards a precise neutrino mass detection from having a very accurate independent determination of $\tau_{\rr{reio}}$, e.g. from surveys focused on reionization and the dark ages,
	\item only a 2- to 3-$\sigma$ detection could be achieved by LSS experiments like Euclid and SKA when combined only with CMB-S4. Indeed, it is important to keep in mind that adding information from low-$\ell$ polarization data strongly constrains $\tau_{\rm reio}$, which leads to a great improvement on the sensitivity to $M_{\nu}$, and therefore CMB-S4 provides much better sensitivity once LiteBIRD,  CORE-M5, or the $\tau_{\rm reio}$ prior is included (similarly, low-$\ell$ Planck data would already help in this regard). This effect is illustrated in \cref{fig:cmb-s4_mnu-tau} (for a physical discussion see~\cite{Archidiacono:2016lnv}).
	\item combining CMB-S4 with either LiteBIRD or CORE-M5 provides similar results in these forecasts. This is due to one of the assumptions made: we assume perfect removal of foregrounds. For a satellite mission with a large number of channels spanning a wide frequency range this is expected to be a reasonable assumption. However, for ground-based missions \textit{without} an accompanying satellite of similar sensitivity and resolution it is less clear if that is so. As such, the CMB-S4 + LiteBIRD results should be viewed as \textit{optimistic}: a more realistic result would be obtained if an additional uncertainty due to foreground cleaning was added. In a similar vein, we can view the CMB-S4 + CORE-M5 as \textit{conservative}: the frequency coverage and large number of channels of the satellite mission would lead to increased faith in foreground cleaning; we do not include the high multipoles for lensing $3000 < \ell < 5000$ for CMB-S4, which would likely lead to improved sensitivity of the combinations involving CMB-S4.
\end{itemize}

\noindent As expected, the neutrino mass sum sensitivity degrades in extended models. Compared to the minimal cosmological model, we notice that in extended models:
\begin{itemize}
	\item the sensitivity only degrades slightly when also varying the number of extra relativistic degrees of freedom, $N_{\rm eff}$, as expected for current and future surveys. The exception is for forecasts involving the $\tau_{\rm reio}$ prior. As we have already seen, if we could precisely determine the optical depth to reionization, we would be able to strongly constrain the neutrino mass sum (at better than 5-$\sigma$, even with current CMB data in combination with Euclid and SKA1 IM). However, when $N_{\rm eff}$ is varying, the $\tau_{\rm reio}$ prior is a little less helpful in precisely measuring the neutrino mass sum, due to non-trivial parameter degeneracies,
	\item the worst sensitivity is always obtained when including a time-varying dark energy equation of state, i.e. in order to obtain a reasonable (3-$\sigma$) level of significance for the detection, CMB plus DESI BAO is never accurate enough and we need at least the combination of LiteBIRD + CMB-S4 or CORE-M5 or PICO together with Euclid and SKA,
	\item however, if we are able to make an accurate, independent measurement of the optical depth to reionization, CORE-M5 or PICO in combination with Euclid and SKA1 intensity mapping would achieve a 4-$\sigma$ neutrino mass detection in {\it any} of the extended models considered.
\end{itemize}

About the extra parameters, we notice that the constraints on $N_{\rm{eff}}$ greatly benefit from the fantastic resolution of CMB-S4 measurements ($\sigma(N_{\rm{eff}})=0.042$ for CMB-S4 alone). A significant improvement with respect to CMB-only constraints is caused by the inclusion of Euclid. Adding SKA intensity mapping does not improve the CMB + Euclid constraints on $N_{\rm eff}$. At the same time, while the $\tau_{\rm reio}$ prior was essential to a robust neutrino mass detection, the constraints on $N_{\rm eff}$ are not sensitive to independent measurements of $\tau_{\rm reio}$, indicating that there is no correlation between the two parameters. The best sensitivity $\sigma(N_{\rm{eff}})=0.017$ is obtained with PICO + Euclid,  while the combination of CMB-S4 + CORE-M5 would yield $\sigma(N_{\rm{eff}})=0.018$. With either CMB-S4 + Euclid or CORE-M5 + Euclid the result is $\sigma(N_{\rm{eff}}) \sim 0.021$, which is more than a factor two improvement with respect to $\sigma(N_{\rm{eff}}) \sim 0.050$ of Planck + Euclid. These results have important consequences for particle physics: it might exclude the existence of any extra relativistic particle, even if the decoupling occurs well before quark confinement~\cite{Baumann:2017gkg}.

For the dark energy equation of state parameters, including large scale structure measurements is crucial. The combination of CMB + Euclid + SKA provides the best sensitivity:  $\sigma(w_0)=0.0022$ and $\lbrace\sigma(w_0)=0.0022, \, \sigma(w_a)=0.017\rbrace$ for Planck + Euclid + SKA. When Euclid and SKA are included, replacing current Planck CMB measurements with a future CMB experiment does not lead to any significant improvement.
Finally, as in the case of $N_{\rm eff}$, applying the $\tau_{\rm reio}$ prior has no impact on the sensitivity.

%% file: conclusions.tex
\section{Conclusions\label{Conclusions}}

In the next decade upcoming large scale structure surveys and future CMB experiments will open up for a detection of a non-zero neutrino mass from cosmology.
In this paper we have presented a forecast of the sensitivity of an extended array of different CMB experiments in conjunction with large-scale structure surveys and external probes of the optical depth at reionization. All forecasts are performed using the same methodology and assumptions, thereby making comparison between similar experiments more accessible.
Allowing for differences in methodology, our results are consistent with previous works when considering similar combinations of experiments and models, e.g.~\cite{Allison:2015qca,Liu:2015txa,Abazajian:2016yjj,Calabrese:2016eii,Lorenz:2017iez,Mishra-Sharma:2018ykh,Yu:2018tem} for CMB-S4,~\cite{DiValentino:2016foa} for CORE-M5,~\cite{Audren:2012vy,Archidiacono:2016lnv,Boyle:2017lzt,Sprenger:2018tdb} for Euclid,~\cite{Villaescusa-Navarro:2015cca,Obuljen:2017jiy,Boyle:2018rva} for SKA, and e.g.~\cite{Font-Ribera:2013rwa} for DESI. We demonstrate the extraordinary complementarity of different cosmological probes, the physical effects of which were studied in detail in e.g.~\cite{Archidiacono:2016lnv,Boyle:2017lzt,Yu:2018tem,Boyle:2018rva}. However, after presenting such encouraging results, we should stress that sensitivity forecasts are always performed with a number of assumptions.

For instance, we make an assumption that the noise in the TE cross-correlation channel is negligible, which may prove to be wrong. Additionally, our assumption that CMB foregrounds can be cleaned up to $\ell_\mathrm{max} = 3000$ for CMB-S4 and PICO might turn out to be too optimistic (for other experiments, the assumption on $\ell_\mathrm{max}$ was less relevant, since the noise spectrum increases exponentially before reaching $\ell_\mathrm{max}$). In particular, for ground-based experiments robustly accounting for foregrounds and systematics with the limitations in sky coverage and number of accessible frequency channels would likely prove very challenging without a satellite of similar sensitivity, with a wide array of channels and full-sky observations. On the other hand, the large size of ground-based telescopes, which are unrealistic targets to reach for satellite missions, allow for resolving very small scales. Therefore, we note that the combination of future high resolution satellite missions and ground-based experiments is crucial to obtain extremely high precision CMB measurements on a wide range of scales.

In the case of Large Scale Structure surveys, we conservatively removed information from non-linear scales, either by marginalising over a theoretical error function growing with $k$ in the case of galaxy surveys, or by removing data above a certain $k$ value recommended by the experimental collaborations in the case of cosmic shear surveys.
We also marginalised over nuisance parameters accounting for uncertainties in the modelling of systematics (e.g. in the galaxy bias function). Despite of all these efforts, additional obstacles might show up in the analysis of future real data (e.g., baryonic feedback effects).

On the side of underlying model assumptions, we focused on a few representative extensions of the $\Lambda$CDM model, but we cannot claim fully model-independent results. In principle, there would always be a possibility to assume a very special modification of General Relativity that would counteract the effect of neutrino masses and relax the bounds (see e.g.~\cite{Hagstotz:2018onp,Hagstotz:2019gsv,Wright:2019qhf,Garcia-Farieta:2019hal}). We could also have a process that changes the relic neutrino distribution~\cite{Oldengott:2019lke}, or future data analyses could return a slightly inconsistent picture calling for a new baseline cosmological model, in which neutrino mass effects might be more difficult to constrain. Finally, it is also possible to assume some new physics causing the decay or the annihilation of cosmological neutrinos into lighter or massless particles, in such way that the neutrino mass would remain forever undetectable by cosmological data~\cite{Beacom:2004yd}. 

While none of these scenarios can be excluded, 
from the present forecasts we can conclude that a neutrino mass detection from cosmology in the next decade is very likely, not only under the assumption that our universe is described by the minimal $\Lambda$CDM model, but even when considering simple, well-motivated extensions to this model.

%% file: appendix.tex
\section{Extended Tables\label{appendix}}
\begin{table}[h!]
	\input{tables/planck_table_new.tex}
	\caption{Expected $1\,\sigma$ sensitivity of Planck alone and in combination with DESI, Euclid, Euclid + SKA1 IM, Euclid + SKA1 IM + $\tau_{reio}$ prior.}
	\label{tab:planck}
\end{table}

\begin{table}
	\input{tables/litebird_table_new.tex}
	\caption{Expected $1\,\sigma$ sensitivity of LiteBIRD alone and in combination with DESI, Euclid, Euclid + SKA1 IM, Euclid + SKA1 IM + $\tau_{reio}$ prior.}
	\label{tab:litebird}
\end{table}

\begin{table}
	\input{tables/cmb_s4_table_new.tex}
	\caption{Expected $1\,\sigma$ sensitivity of CMB S4 alone and in combination with DESI, Euclid, Euclid + SKA1 IM, Euclid + SKA1 IM + $\tau_{reio}$ prior.}
	\label{tab:s4}
\end{table}

\begin{table}
	\input{tables/cmb_s4_litebird_table_new.tex}
	\caption{Expected $1\,\sigma$ sensitivity of CMB S4 + LiteBIRD alone and in combination with DESI, Euclid, Euclid + SKA1 IM, Euclid + SKA1 IM + $\tau_{reio}$ prior.}
	\label{tab:s4+litebird}
\end{table}

\begin{table}
	\input{tables/core_table_new.tex}
	\caption{Expected $1\,\sigma$ sensitivity of CORE-M5 alone and in combination with DESI, Euclid, Euclid + SKA1 IM, Euclid + SKA1 IM + $\tau_{reio}$ prior.}
	\label{tab:core}
\end{table}

\begin{table}
	\input{tables/core_s4_table_new.tex}
	\caption{Expected $1\,\sigma$ sensitivity of CMB S4 + CORE-M5 alone and in combination with DESI, Euclid, Euclid + SKA1 IM, Euclid + SKA1 IM + $\tau_{reio}$ prior.}
	\label{tab:s4+core}
\end{table}

\begin{table}
	\input{tables/pico_table_new.tex}
	\caption{Expected $1\,\sigma$ sensitivity of PICO alone and in combination with DESI, Euclid, Euclid + SKA1 IM, Euclid + SKA1 IM + $\tau_{reio}$ prior.}
	\label{tab:pico}
\end{table}

%% file: tables/planck_table_new.tex
\resizebox{1.0\textwidth}{!}{%
\begingroup
\setlength{\tabcolsep}{10pt}
\renewcommand{\arraystretch}{1.5}
\begin{tabular}{|c|c|c|c|c|c|c|}
\cline{3-7}
\multicolumn{2}{c|}{}&CMB only & CMB + DESI & CMB + Euclid & + SKA & + $\tau_{\mathrm{reio}}$ prior \\
 \hline   & $\sigma \left( 100*\omega_{\mathrm{b}} \right)$ & 0.016 & 0.013 & 0.012 & 0.011 & 0.011\\ 
    & $\sigma \left( \omega_{\mathrm{cdm}} \right)$ & 0.0014 & 0.00076 & 0.00029 & 0.00029 & 0.00018\\ 
   $\Lambda$CDM & $\sigma \left( H_0 \right) / [\frac{\mathrm{km}}{\mathrm{s\,Mpc}}]$ & 1.4 & 0.25 & 0.14 & 0.081 & 0.079\\ 
    & $\sigma \left( \ln 10^{10}A_{s } \right)$ & 0.0089 & 0.0089 & 0.0079 & 0.0076 & 0.0024\\ 
   + $M_{\nu}$ & $\sigma \left( n_s \right)$ & 0.004 & 0.003 & 0.00077 & 0.0007 & 0.00061\\ 
    & $\sigma \left( \tau_{\mathrm{reio}} \right)$ & 0.0045 & 0.0045 & 0.0041 & 0.004 & 0.00098\\ 
    & $\sigma \left(M_{\nu}\right) / [eV]$ & 0.097 & 0.044 & 0.02 & 0.016 & 0.0082\\ 
 \hline\hline   & $\sigma \left( 100*\omega_{\mathrm{b}} \right)$ & 0.025 & 0.019 & 0.013 & 0.013 & 0.013\\ 
    & $\sigma \left( \omega_{\mathrm{cdm}} \right)$ & 0.003 & 0.003 & 0.00087 & 0.00084 & 0.0008\\ 
   $\Lambda$CDM & $\sigma \left( H_0 \right) / [\frac{\mathrm{km}}{\mathrm{s\,Mpc}}]$ & 2.1 & 0.91 & 0.25 & 0.23 & 0.23\\ 
    & $\sigma \left( \ln 10^{10}A_{s } \right)$ & 0.013 & 0.012 & 0.0083 & 0.0077 & 0.0037\\ 
   + $M_{\nu}$ & $\sigma \left( n_s \right)$ & 0.0091 & 0.0069 & 0.00085 & 0.0008 & 0.0007\\ 
    & $\sigma \left( \tau_{\mathrm{reio}} \right)$ & 0.0046 & 0.0045 & 0.0041 & 0.0039 & 0.00099\\ 
    + $N_{\mathrm{eff}}$ & $\sigma \left(M_{\nu}\right) / [eV]$ & 0.099 & 0.047 & 0.023 & 0.018 & 0.014\\ 
    & $\sigma \left( N_{\mathrm{eff}} \right)$ & 0.19 & 0.17 & 0.05 & 0.048 & 0.049\\ 
 \hline\hline   & $\sigma \left( 100*\omega_{\mathrm{b}} \right)$ & 0.017 & 0.013 & 0.012 & 0.011 & 0.011\\ 
    & $\sigma \left( \omega_{\mathrm{cdm}} \right)$ & 0.0016 & 0.00082 & 0.0003 & 0.00028 & 0.00019\\ 
   CDM & $\sigma \left( H_0 \right) / [\frac{\mathrm{km}}{\mathrm{s\,Mpc}}]$ & 21 & 0.84 & 0.24 & 0.087 & 0.082\\ 
    & $\sigma \left( \ln 10^{10}A_{s } \right)$ & 0.0089 & 0.0089 & 0.008 & 0.0077 & 0.0024\\ 
   + $M_{\nu}$ & $\sigma \left( n_s \right)$ & 0.0043 & 0.0031 & 0.00098 & 0.00071 & 0.0006\\ 
    & $\sigma \left( \tau_{\mathrm{reio}} \right)$ & 0.0047 & 0.0045 & 0.0042 & 0.0041 & 0.001\\ 
   + $w_{0}$ & $\sigma \left(M_{\nu}\right) / [eV]$ & 0.14 & 0.051 & 0.027 & 0.019 & 0.011\\ 
    & $\sigma \left( w_{0} \right)$ & 0.48 & 0.038 & 0.012 & 0.0022 & 0.0022\\ 
 \hline\hline   & $\sigma \left( 100*\omega_{\mathrm{b}} \right)$ & 0.018 & 0.013 & 0.011 & 0.011 & 0.011\\ 
   CDM & $\sigma \left( \omega_{\mathrm{cdm}} \right)$ & 0.0016 & 0.00092 & 0.0003 & 0.00029 & 0.00021\\ 
    & $\sigma \left( H_0 \right) / [\frac{\mathrm{km}}{\mathrm{s\,Mpc}}]$ & 20 & 1.7 & 0.27 & 0.095 & 0.087\\ 
   + $M_{\nu}$ & $\sigma \left( \ln 10^{10}A_{s } \right)$ & 0.009 & 0.0087 & 0.0077 & 0.0078 & 0.0024\\ 
    & $\sigma \left( n_s \right)$ & 0.0043 & 0.0032 & 0.00095 & 0.00075 & 0.00062\\ 
   + $w_{0}$ & $\sigma \left( \tau_{\mathrm{reio}} \right)$ & 0.0046 & 0.0045 & 0.0041 & 0.0041 & 0.00098\\ 
    & $\sigma \left(M_{\nu}\right) / [eV]$ & 0.14 & 0.069 & 0.03 & 0.027 & 0.017\\ 
   + $w_{a}$ & $\sigma \left( w_{0} \right)$ & 0.84 & 0.2 & 0.021 & 0.0022 & 0.0021\\ 
    & $\sigma \left( w_{a} \right)$ & 2.2 & 0.52 & 0.064 & 0.017 & 0.016\\ 
 \hline\end{tabular} 
 \endgroup}

%% file: tables/litebird_table_new.tex
\resizebox{1.0\textwidth}{!}{%
\begingroup
\setlength{\tabcolsep}{10pt}
\renewcommand{\arraystretch}{1.5}
\begin{tabular}{|c|c|c|c|c|c|c|}
\cline{3-7}
\multicolumn{2}{c|}{}&CMB only & CMB + DESI & CMB + Euclid & + SKA & + $\tau_{\mathrm{reio}}$ prior \\
 \hline   & $\sigma \left( 100*\omega_{\mathrm{b}} \right)$ & 0.018 & 0.014 & 0.011 & 0.011 & 0.011\\ 
    & $\sigma \left( \omega_{\mathrm{cdm}} \right)$ & 0.0011 & 0.00072 & 0.00027 & 0.00025 & 0.00023\\ 
   $\Lambda$CDM & $\sigma \left( H_0 \right) / [\frac{\mathrm{km}}{\mathrm{s\,Mpc}}]$ & 1.3 & 0.25 & 0.13 & 0.089 & 0.088\\ 
    & $\sigma \left( \ln 10^{10}A_{s } \right)$ & 0.0051 & 0.0052 & 0.0045 & 0.0044 & 0.0028\\ 
   + $M_{\nu}$ & $\sigma \left( n_s \right)$ & 0.0045 & 0.004 & 0.00074 & 0.00067 & 0.00065\\ 
    & $\sigma \left( \tau_{\mathrm{reio}} \right)$ & 0.0022 & 0.0022 & 0.0021 & 0.002 & 0.0009\\ 
    & $\sigma \left(M_{\nu}\right) / [eV]$ & 0.095 & 0.044 & 0.016 & 0.012 & 0.01\\ 
 \hline\hline   & $\sigma \left( 100*\omega_{\mathrm{b}} \right)$ & 0.025 & 0.02 & 0.012 & 0.012 & 0.012\\ 
    & $\sigma \left( \omega_{\mathrm{cdm}} \right)$ & 0.0046 & 0.0045 & 0.0013 & 0.0012 & 0.0012\\ 
   $\Lambda$CDM & $\sigma \left( H_0 \right) / [\frac{\mathrm{km}}{\mathrm{s\,Mpc}}]$ & 1.9 & 1.3 & 0.34 & 0.31 & 0.32\\ 
    & $\sigma \left( \ln 10^{10}A_{s } \right)$ & 0.012 & 0.012 & 0.005 & 0.0048 & 0.0035\\ 
   + $M_{\nu}$ & $\sigma \left( n_s \right)$ & 0.008 & 0.0075 & 0.001 & 0.001 & 0.00099\\ 
    & $\sigma \left( \tau_{\mathrm{reio}} \right)$ & 0.0022 & 0.0021 & 0.0021 & 0.0021 & 0.00093\\ 
    + $N_{\mathrm{eff}}$ & $\sigma \left(M_{\nu}\right) / [eV]$ & 0.098 & 0.047 & 0.021 & 0.017 & 0.016\\ 
    & $\sigma \left( N_{\mathrm{eff}} \right)$ & 0.27 & 0.26 & 0.078 & 0.074 & 0.075\\ 
 \hline\hline   & $\sigma \left( 100*\omega_{\mathrm{b}} \right)$ & 0.018 & 0.015 & 0.011 & 0.011 & 0.011\\ 
    & $\sigma \left( \omega_{\mathrm{cdm}} \right)$ & 0.0012 & 0.00078 & 0.00027 & 0.00025 & 0.00023\\ 
   CDM & $\sigma \left( H_0 \right) / [\frac{\mathrm{km}}{\mathrm{s\,Mpc}}]$ & 21 & 0.85 & 0.25 & 0.091 & 0.093\\ 
    & $\sigma \left( \ln 10^{10}A_{s } \right)$ & 0.0052 & 0.0051 & 0.0046 & 0.0045 & 0.0029\\ 
   + $M_{\nu}$ & $\sigma \left( n_s \right)$ & 0.0046 & 0.004 & 0.00097 & 0.00067 & 0.00066\\ 
    & $\sigma \left( \tau_{\mathrm{reio}} \right)$ & 0.0022 & 0.0021 & 0.0021 & 0.0021 & 0.0009\\ 
   + $w_{0}$ & $\sigma \left(M_{\nu}\right) / [eV]$ & 0.13 & 0.051 & 0.024 & 0.015 & 0.013\\ 
    & $\sigma \left( w_{0} \right)$ & 0.48 & 0.038 & 0.012 & 0.0022 & 0.0021\\ 
 \hline\hline   & $\sigma \left( 100*\omega_{\mathrm{b}} \right)$ & 0.019 & 0.015 & 0.011 & 0.011 & 0.011\\ 
   CDM & $\sigma \left( \omega_{\mathrm{cdm}} \right)$ & 0.0012 & 0.00083 & 0.00028 & 0.00028 & 0.00026\\ 
    & $\sigma \left( H_0 \right) / [\frac{\mathrm{km}}{\mathrm{s\,Mpc}}]$ & 20 & 1.7 & 0.27 & 0.096 & 0.097\\ 
   + $M_{\nu}$ & $\sigma \left( \ln 10^{10}A_{s } \right)$ & 0.0052 & 0.0051 & 0.0046 & 0.0046 & 0.0029\\ 
    & $\sigma \left( n_s \right)$ & 0.0047 & 0.0039 & 0.00097 & 0.0007 & 0.00067\\ 
   + $w_{0}$ & $\sigma \left( \tau_{\mathrm{reio}} \right)$ & 0.0023 & 0.0022 & 0.0021 & 0.0021 & 0.0009\\ 
    & $\sigma \left(M_{\nu}\right) / [eV]$ & 0.13 & 0.068 & 0.031 & 0.022 & 0.02\\ 
   + $w_{a}$ & $\sigma \left( w_{0} \right)$ & 0.81 & 0.19 & 0.021 & 0.0022 & 0.0021\\ 
    & $\sigma \left( w_{a} \right)$ & 2.1 & 0.5 & 0.068 & 0.017 & 0.017\\ 
 \hline\end{tabular} 
 \endgroup}

%% file: tables/cmb_s4_table_new.tex
\resizebox{1.0\textwidth}{!}{%
\begingroup
\setlength{\tabcolsep}{10pt}
\renewcommand{\arraystretch}{1.5}
\begin{tabular}{|c|c|c|c|c|c|c|}
\cline{3-7}
\multicolumn{2}{c|}{}&CMB only & CMB + DESI & CMB + Euclid & + SKA & + $\tau_{\mathrm{reio}}$ prior \\
 \hline   & $\sigma \left( 100*\omega_{\mathrm{b}} \right)$ & 0.0035 & 0.0034 & 0.0026 & 0.0026 & 0.0026\\ 
    & $\sigma \left( \omega_{\mathrm{cdm}} \right)$ & 0.00079 & 0.00064 & 0.00043 & 0.0004 & 8.9e-05\\ 
   $\Lambda$CDM & $\sigma \left( H_0 \right) / [\frac{\mathrm{km}}{\mathrm{s\,Mpc}}]$ & 0.77 & 0.24 & 0.15 & 0.063 & 0.041\\ 
    & $\sigma \left( \ln 10^{10}A_{s } \right)$ & 0.022 & 0.017 & 0.014 & 0.012 & 0.002\\ 
   + $M_{\nu}$ & $\sigma \left( n_s \right)$ & 0.0024 & 0.0022 & 0.00098 & 0.00084 & 0.00056\\ 
    & $\sigma \left( \tau_{\mathrm{reio}} \right)$ & 0.011 & 0.0096 & 0.0074 & 0.0065 & 0.001\\ 
    & $\sigma \left(M_{\nu}\right) / [eV]$ & 0.075 & 0.04 & 0.032 & 0.023 & 0.006\\ 
 \hline\hline   & $\sigma \left( 100*\omega_{\mathrm{b}} \right)$ & 0.0051 & 0.005 & 0.0046 & 0.0046 & 0.0046\\ 
    & $\sigma \left( \omega_{\mathrm{cdm}} \right)$ & 0.00092 & 0.00078 & 0.00057 & 0.00053 & 0.00036\\ 
   $\Lambda$CDM & $\sigma \left( H_0 \right) / [\frac{\mathrm{km}}{\mathrm{s\,Mpc}}]$ & 0.85 & 0.29 & 0.18 & 0.12 & 0.11\\ 
    & $\sigma \left( \ln 10^{10}A_{s } \right)$ & 0.024 & 0.018 & 0.014 & 0.012 & 0.0026\\ 
   + $M_{\nu}$ & $\sigma \left( n_s \right)$ & 0.0039 & 0.0034 & 0.00097 & 0.00084 & 0.00057\\ 
    & $\sigma \left( \tau_{\mathrm{reio}} \right)$ & 0.012 & 0.0099 & 0.0074 & 0.0065 & 0.001\\ 
    + $N_{\mathrm{eff}}$ & $\sigma \left(M_{\nu}\right) / [eV]$ & 0.081 & 0.043 & 0.033 & 0.024 & 0.0082\\ 
    & $\sigma \left( N_{\mathrm{eff}} \right)$ & 0.042 & 0.039 & 0.022 & 0.021 & 0.021\\ 
 \hline\hline   & $\sigma \left( 100*\omega_{\mathrm{b}} \right)$ & 0.0034 & 0.0035 & 0.0027 & 0.0026 & 0.0026\\ 
    & $\sigma \left( \omega_{\mathrm{cdm}} \right)$ & 0.00086 & 0.00071 & 0.00043 & 0.0004 & 0.00011\\ 
   CDM & $\sigma \left( H_0 \right) / [\frac{\mathrm{km}}{\mathrm{s\,Mpc}}]$ & 10 & 0.83 & 0.22 & 0.076 & 0.048\\ 
    & $\sigma \left( \ln 10^{10}A_{s } \right)$ & 0.039 & 0.018 & 0.014 & 0.012 & 0.002\\ 
   + $M_{\nu}$ & $\sigma \left( n_s \right)$ & 0.0024 & 0.0022 & 0.0011 & 0.00087 & 0.00056\\ 
    & $\sigma \left( \tau_{\mathrm{reio}} \right)$ & 0.02 & 0.01 & 0.0077 & 0.0068 & 0.001\\ 
   + $w_{0}$ & $\sigma \left(M_{\nu}\right) / [eV]$ & 0.08 & 0.046 & 0.037 & 0.028 & 0.0093\\ 
    & $\sigma \left( w_{0} \right)$ & 0.18 & 0.036 & 0.01 & 0.0023 & 0.0021\\ 
 \hline\hline   & $\sigma \left( 100*\omega_{\mathrm{b}} \right)$ & 0.0034 & 0.0035 & 0.0027 & 0.0026 & 0.0027\\ 
   CDM & $\sigma \left( \omega_{\mathrm{cdm}} \right)$ & 0.00086 & 0.00077 & 0.00042 & 0.00038 & 0.00014\\ 
    & $\sigma \left( H_0 \right) / [\frac{\mathrm{km}}{\mathrm{s\,Mpc}}]$ & 12 & 1.6 & 0.27 & 0.089 & 0.055\\ 
   + $M_{\nu}$ & $\sigma \left( \ln 10^{10}A_{s } \right)$ & 0.036 & 0.02 & 0.014 & 0.013 & 0.0021\\ 
    & $\sigma \left( n_s \right)$ & 0.0025 & 0.0024 & 0.0011 & 0.00092 & 0.00057\\ 
   + $w_{0}$ & $\sigma \left( \tau_{\mathrm{reio}} \right)$ & 0.019 & 0.011 & 0.0076 & 0.007 & 0.001\\ 
    & $\sigma \left(M_{\nu}\right) / [eV]$ & 0.076 & 0.062 & 0.042 & 0.037 & 0.016\\ 
   + $w_{a}$ & $\sigma \left( w_{0} \right)$ & 0.37 & 0.18 & 0.021 & 0.0022 & 0.0021\\ 
    & $\sigma \left( w_{a} \right)$ & 2.1 & 0.38 & 0.069 & 0.017 & 0.016\\ 
 \hline\end{tabular} 
 \endgroup}

%% file: tables/cmb_s4_litebird_table_new.tex
\resizebox{1.0\textwidth}{!}{%
\begingroup
\setlength{\tabcolsep}{10pt}
\renewcommand{\arraystretch}{1.5}
\begin{tabular}{|c|c|c|c|c|c|c|}
\cline{3-7}
\multicolumn{2}{c|}{}&CMB only & CMB + DESI & CMB + Euclid & + SKA & + $\tau_{\mathrm{reio}}$ prior \\
 \hline   & $\sigma \left( 100*\omega_{\mathrm{b}} \right)$ & 0.0033 & 0.0032 & 0.0026 & 0.0025 & 0.0025\\ 
    & $\sigma \left( \omega_{\mathrm{cdm}} \right)$ & 0.00053 & 0.00025 & 0.00017 & 0.00013 & 8.4e-05\\ 
   $\Lambda$CDM & $\sigma \left( H_0 \right) / [\frac{\mathrm{km}}{\mathrm{s\,Mpc}}]$ & 0.54 & 0.23 & 0.11 & 0.042 & 0.039\\ 
    & $\sigma \left( \ln 10^{10}A_{s } \right)$ & 0.004 & 0.0039 & 0.0039 & 0.0037 & 0.0018\\ 
   + $M_{\nu}$ & $\sigma \left( n_s \right)$ & 0.0018 & 0.0015 & 0.00063 & 0.00058 & 0.00055\\ 
    & $\sigma \left( \tau_{\mathrm{reio}} \right)$ & 0.002 & 0.002 & 0.0021 & 0.002 & 0.0009\\ 
    & $\sigma \left(M_{\nu}\right) / [eV]$ & 0.042 & 0.019 & 0.012 & 0.0084 & 0.0056\\ 
 \hline\hline   & $\sigma \left( 100*\omega_{\mathrm{b}} \right)$ & 0.005 & 0.0048 & 0.0044 & 0.0043 & 0.0044\\ 
    & $\sigma \left( \omega_{\mathrm{cdm}} \right)$ & 0.00072 & 0.00061 & 0.00038 & 0.00035 & 0.00034\\ 
   $\Lambda$CDM & $\sigma \left( H_0 \right) / [\frac{\mathrm{km}}{\mathrm{s\,Mpc}}]$ & 0.63 & 0.29 & 0.14 & 0.1 & 0.1\\ 
    & $\sigma \left( \ln 10^{10}A_{s } \right)$ & 0.0044 & 0.0043 & 0.0041 & 0.004 & 0.0023\\ 
   + $M_{\nu}$ & $\sigma \left( n_s \right)$ & 0.0029 & 0.0026 & 0.00062 & 0.00057 & 0.00055\\ 
    & $\sigma \left( \tau_{\mathrm{reio}} \right)$ & 0.0021 & 0.0021 & 0.002 & 0.002 & 0.00091\\ 
    + $N_{\mathrm{eff}}$ & $\sigma \left(M_{\nu}\right) / [eV]$ & 0.042 & 0.021 & 0.014 & 0.0098 & 0.0078\\ 
    & $\sigma \left( N_{\mathrm{eff}} \right)$ & 0.038 & 0.037 & 0.021 & 0.02 & 0.02\\ 
 \hline\hline   & $\sigma \left( 100*\omega_{\mathrm{b}} \right)$ & 0.0033 & 0.0033 & 0.0026 & 0.0025 & 0.0025\\ 
    & $\sigma \left( \omega_{\mathrm{cdm}} \right)$ & 0.00061 & 0.00032 & 0.00017 & 0.00015 & 0.0001\\ 
   CDM & $\sigma \left( H_0 \right) / [\frac{\mathrm{km}}{\mathrm{s\,Mpc}}]$ & 6.2 & 0.82 & 0.21 & 0.051 & 0.047\\ 
    & $\sigma \left( \ln 10^{10}A_{s } \right)$ & 0.0042 & 0.0039 & 0.0038 & 0.0038 & 0.0018\\ 
   + $M_{\nu}$ & $\sigma \left( n_s \right)$ & 0.002 & 0.0016 & 0.00074 & 0.00059 & 0.00055\\ 
    & $\sigma \left( \tau_{\mathrm{reio}} \right)$ & 0.0021 & 0.0021 & 0.0021 & 0.002 & 0.00091\\ 
   + $w_{0}$ & $\sigma \left(M_{\nu}\right) / [eV]$ & 0.063 & 0.034 & 0.017 & 0.012 & 0.0088\\ 
    & $\sigma \left( w_{0} \right)$ & 0.11 & 0.032 & 0.011 & 0.0022 & 0.0021\\ 
 \hline\hline   & $\sigma \left( 100*\omega_{\mathrm{b}} \right)$ & 0.0033 & 0.0033 & 0.0026 & 0.0026 & 0.0026\\ 
   CDM & $\sigma \left( \omega_{\mathrm{cdm}} \right)$ & 0.00063 & 0.00039 & 0.00018 & 0.00016 & 0.00014\\ 
    & $\sigma \left( H_0 \right) / [\frac{\mathrm{km}}{\mathrm{s\,Mpc}}]$ & 8.9 & 1.6 & 0.26 & 0.059 & 0.054\\ 
   + $M_{\nu}$ & $\sigma \left( \ln 10^{10}A_{s } \right)$ & 0.0043 & 0.0039 & 0.0039 & 0.0039 & 0.0019\\ 
    & $\sigma \left( n_s \right)$ & 0.002 & 0.0016 & 0.00074 & 0.00059 & 0.00055\\ 
   + $w_{0}$ & $\sigma \left( \tau_{\mathrm{reio}} \right)$ & 0.002 & 0.0021 & 0.0021 & 0.002 & 0.00093\\ 
    & $\sigma \left(M_{\nu}\right) / [eV]$ & 0.063 & 0.051 & 0.023 & 0.018 & 0.015\\ 
   + $w_{a}$ & $\sigma \left( w_{0} \right)$ & 0.31 & 0.18 & 0.02 & 0.0022 & 0.0022\\ 
    & $\sigma \left( w_{a} \right)$ & 1.8 & 0.45 & 0.068 & 0.017 & 0.016\\ 
 \hline\end{tabular} 
 \endgroup}

%% file: tables/core_table_new.tex
\resizebox{1.0\textwidth}{!}{%
\begingroup
\setlength{\tabcolsep}{10pt}
\renewcommand{\arraystretch}{1.5}
\begin{tabular}{|c|c|c|c|c|c|c|}
\cline{3-7}
\multicolumn{2}{c|}{}&CMB only & CMB + DESI & CMB + Euclid & + SKA & + $\tau_{\mathrm{reio}}$ prior \\
 \hline   & $\sigma \left( 100*\omega_{\mathrm{b}} \right)$ & 0.0039 & 0.0039 & 0.0029 & 0.0029 & 0.0029\\ 
    & $\sigma \left( \omega_{\mathrm{cdm}} \right)$ & 0.00053 & 0.00025 & 0.00016 & 0.00013 & 8.2e-05\\ 
   $\Lambda$CDM & $\sigma \left( H_0 \right) / [\frac{\mathrm{km}}{\mathrm{s\,Mpc}}]$ & 0.54 & 0.22 & 0.11 & 0.042 & 0.04\\ 
    & $\sigma \left( \ln 10^{10}A_{s } \right)$ & 0.004 & 0.0039 & 0.0039 & 0.0036 & 0.0018\\ 
   + $M_{\nu}$ & $\sigma \left( n_s \right)$ & 0.0017 & 0.0015 & 0.00062 & 0.00058 & 0.00055\\ 
    & $\sigma \left( \tau_{\mathrm{reio}} \right)$ & 0.0021 & 0.002 & 0.0021 & 0.0019 & 0.00089\\ 
    & $\sigma \left(M_{\nu}\right) / [eV]$ & 0.042 & 0.02 & 0.012 & 0.0083 & 0.0056\\ 
 \hline\hline   & $\sigma \left( 100*\omega_{\mathrm{b}} \right)$ & 0.0059 & 0.0057 & 0.005 & 0.0052 & 0.0051\\ 
    & $\sigma \left( \omega_{\mathrm{cdm}} \right)$ & 0.00073 & 0.00067 & 0.00039 & 0.00037 & 0.00036\\ 
   $\Lambda$CDM & $\sigma \left( H_0 \right) / [\frac{\mathrm{km}}{\mathrm{s\,Mpc}}]$ & 0.64 & 0.31 & 0.14 & 0.11 & 0.11\\ 
    & $\sigma \left( \ln 10^{10}A_{s } \right)$ & 0.0044 & 0.0043 & 0.004 & 0.004 & 0.0024\\ 
   + $M_{\nu}$ & $\sigma \left( n_s \right)$ & 0.003 & 0.0026 & 0.00062 & 0.00057 & 0.00055\\ 
    & $\sigma \left( \tau_{\mathrm{reio}} \right)$ & 0.0021 & 0.0021 & 0.002 & 0.002 & 0.00091\\ 
    + $N_{\mathrm{eff}}$ & $\sigma \left(M_{\nu}\right) / [eV]$ & 0.042 & 0.021 & 0.014 & 0.0099 & 0.0078\\ 
    & $\sigma \left( N_{\mathrm{eff}} \right)$ & 0.041 & 0.04 & 0.021 & 0.021 & 0.021\\ 
 \hline\hline   & $\sigma \left( 100*\omega_{\mathrm{b}} \right)$ & 0.0039 & 0.0037 & 0.0031 & 0.0029 & 0.0029\\ 
    & $\sigma \left( \omega_{\mathrm{cdm}} \right)$ & 0.00058 & 0.00031 & 0.00017 & 0.00014 & 0.0001\\ 
   CDM & $\sigma \left( H_0 \right) / [\frac{\mathrm{km}}{\mathrm{s\,Mpc}}]$ & 5.2 & 0.8 & 0.21 & 0.05 & 0.048\\ 
    & $\sigma \left( \ln 10^{10}A_{s } \right)$ & 0.0041 & 0.0038 & 0.0037 & 0.0037 & 0.0018\\ 
   + $M_{\nu}$ & $\sigma \left( n_s \right)$ & 0.0018 & 0.0015 & 0.00073 & 0.00058 & 0.00053\\ 
    & $\sigma \left( \tau_{\mathrm{reio}} \right)$ & 0.002 & 0.0021 & 0.002 & 0.002 & 0.00091\\ 
   + $w_{0}$ & $\sigma \left(M_{\nu}\right) / [eV]$ & 0.058 & 0.034 & 0.017 & 0.011 & 0.009\\ 
    & $\sigma \left( w_{0} \right)$ & 0.12 & 0.032 & 0.01 & 0.0021 & 0.0021\\ 
 \hline\hline   & $\sigma \left( 100*\omega_{\mathrm{b}} \right)$ & 0.004 & 0.0038 & 0.003 & 0.003 & 0.003\\ 
   CDM & $\sigma \left( \omega_{\mathrm{cdm}} \right)$ & 0.00061 & 0.00037 & 0.00018 & 0.00016 & 0.00014\\ 
    & $\sigma \left( H_0 \right) / [\frac{\mathrm{km}}{\mathrm{s\,Mpc}}]$ & 8.2 & 1.6 & 0.25 & 0.058 & 0.056\\ 
   + $M_{\nu}$ & $\sigma \left( \ln 10^{10}A_{s } \right)$ & 0.0042 & 0.0039 & 0.0037 & 0.0037 & 0.0018\\ 
    & $\sigma \left( n_s \right)$ & 0.0019 & 0.0015 & 0.00072 & 0.0006 & 0.00056\\ 
   + $w_{0}$ & $\sigma \left( \tau_{\mathrm{reio}} \right)$ & 0.002 & 0.0021 & 0.002 & 0.002 & 0.00091\\ 
    & $\sigma \left(M_{\nu}\right) / [eV]$ & 0.059 & 0.049 & 0.023 & 0.017 & 0.016\\ 
   + $w_{a}$ & $\sigma \left( w_{0} \right)$ & 0.3 & 0.18 & 0.02 & 0.0021 & 0.0022\\ 
    & $\sigma \left( w_{a} \right)$ & 1.9 & 0.43 & 0.068 & 0.016 & 0.017\\ 
 \hline\end{tabular} 
 \endgroup}

%% file: tables/core_s4_table_new.tex
\resizebox{1.0\textwidth}{!}{%
\begingroup
\setlength{\tabcolsep}{10pt}
\renewcommand{\arraystretch}{1.5}
\begin{tabular}{|c|c|c|c|c|c|c|}
\cline{3-7}
\multicolumn{2}{c|}{}&CMB only & CMB + DESI & CMB + Euclid & + SKA & + $\tau_{\mathrm{reio}}$ prior \\
 \hline   & $\sigma \left( 100*\omega_{\mathrm{b}} \right)$ & 0.0029 & 0.0029 & 0.0023 & 0.0022 & 0.0022\\ 
    & $\sigma \left( \omega_{\mathrm{cdm}} \right)$ & 0.0005 & 0.00025 & 0.00016 & 0.00013 & 8e-05\\ 
   $\Lambda$CDM & $\sigma \left( H_0 \right) / [\frac{\mathrm{km}}{\mathrm{s\,Mpc}}]$ & 0.51 & 0.22 & 0.11 & 0.041 & 0.039\\ 
    & $\sigma \left( \ln 10^{10}A_{s } \right)$ & 0.004 & 0.0039 & 0.0037 & 0.0037 & 0.0018\\ 
   + $M_{\nu}$ & $\sigma \left( n_s \right)$ & 0.0017 & 0.0014 & 0.00061 & 0.00057 & 0.00055\\ 
    & $\sigma \left( \tau_{\mathrm{reio}} \right)$ & 0.002 & 0.0021 & 0.002 & 0.002 & 0.00091\\ 
    & $\sigma \left(M_{\nu}\right) / [eV]$ & 0.041 & 0.019 & 0.012 & 0.0084 & 0.0054\\ 
 \hline\hline   & $\sigma \left( 100*\omega_{\mathrm{b}} \right)$ & 0.0043 & 0.0043 & 0.0039 & 0.0039 & 0.0039\\ 
    & $\sigma \left( \omega_{\mathrm{cdm}} \right)$ & 0.00065 & 0.00055 & 0.00034 & 0.00032 & 0.0003\\ 
   $\Lambda$CDM & $\sigma \left( H_0 \right) / [\frac{\mathrm{km}}{\mathrm{s\,Mpc}}]$ & 0.59 & 0.28 & 0.13 & 0.093 & 0.093\\ 
    & $\sigma \left( \ln 10^{10}A_{s } \right)$ & 0.0041 & 0.0041 & 0.004 & 0.0038 & 0.0022\\ 
   + $M_{\nu}$ & $\sigma \left( n_s \right)$ & 0.0027 & 0.0023 & 0.00061 & 0.00057 & 0.00053\\ 
    & $\sigma \left( \tau_{\mathrm{reio}} \right)$ & 0.002 & 0.002 & 0.002 & 0.002 & 0.00091\\ 
    + $N_{\mathrm{eff}}$ & $\sigma \left(M_{\nu}\right) / [eV]$ & 0.041 & 0.02 & 0.013 & 0.0092 & 0.0072\\ 
    & $\sigma \left( N_{\mathrm{eff}} \right)$ & 0.033 & 0.032 & 0.018 & 0.018 & 0.018\\ 
 \hline\hline   & $\sigma \left( 100*\omega_{\mathrm{b}} \right)$ & 0.003 & 0.0029 & 0.0023 & 0.0023 & 0.0023\\ 
    & $\sigma \left( \omega_{\mathrm{cdm}} \right)$ & 0.00055 & 0.0003 & 0.00017 & 0.00014 & 0.0001\\ 
   CDM & $\sigma \left( H_0 \right) / [\frac{\mathrm{km}}{\mathrm{s\,Mpc}}]$ & 4.6 & 0.8 & 0.21 & 0.049 & 0.046\\ 
    & $\sigma \left( \ln 10^{10}A_{s } \right)$ & 0.0042 & 0.0039 & 0.0038 & 0.0036 & 0.0018\\ 
   + $M_{\nu}$ & $\sigma \left( n_s \right)$ & 0.0018 & 0.0014 & 0.00071 & 0.00057 & 0.00054\\ 
    & $\sigma \left( \tau_{\mathrm{reio}} \right)$ & 0.0021 & 0.0021 & 0.002 & 0.002 & 0.0009\\ 
   + $w_{0}$ & $\sigma \left(M_{\nu}\right) / [eV]$ & 0.056 & 0.034 & 0.017 & 0.011 & 0.0088\\ 
    & $\sigma \left( w_{0} \right)$ & 0.12 & 0.032 & 0.011 & 0.0021 & 0.0021\\ 
 \hline\hline   & $\sigma \left( 100*\omega_{\mathrm{b}} \right)$ & 0.003 & 0.0029 & 0.0024 & 0.0023 & 0.0023\\ 
   CDM & $\sigma \left( \omega_{\mathrm{cdm}} \right)$ & 0.00059 & 0.00035 & 0.00018 & 0.00016 & 0.00014\\ 
    & $\sigma \left( H_0 \right) / [\frac{\mathrm{km}}{\mathrm{s\,Mpc}}]$ & 7.5 & 1.6 & 0.26 & 0.057 & 0.052\\ 
   + $M_{\nu}$ & $\sigma \left( \ln 10^{10}A_{s } \right)$ & 0.0042 & 0.0037 & 0.0037 & 0.0038 & 0.0018\\ 
    & $\sigma \left( n_s \right)$ & 0.0018 & 0.0015 & 0.00072 & 0.00059 & 0.00055\\ 
   + $w_{0}$ & $\sigma \left( \tau_{\mathrm{reio}} \right)$ & 0.002 & 0.0019 & 0.002 & 0.002 & 0.0009\\ 
    & $\sigma \left(M_{\nu}\right) / [eV]$ & 0.056 & 0.035 & 0.023 & 0.017 & 0.015\\ 
   + $w_{a}$ & $\sigma \left( w_{0} \right)$ & 0.29 & 0.17 & 0.021 & 0.0021 & 0.0021\\ 
    & $\sigma \left( w_{a} \right)$ & 1.8 & 0.42 & 0.069 & 0.016 & 0.016\\ 
 \hline\end{tabular} 
 \endgroup}

%% file: tables/pico_table_new.tex
\resizebox{1.0\textwidth}{!}{%
\begingroup
\setlength{\tabcolsep}{10pt}
\renewcommand{\arraystretch}{1.5}
\begin{tabular}{|c|c|c|c|c|c|c|}
\cline{3-7}
\multicolumn{2}{c|}{}&CMB only & CMB + DESI & CMB + Euclid & + SKA & + $\tau_{\mathrm{reio}}$ prior \\
 \hline   & $\sigma \left( 100*\omega_{\mathrm{b}} \right)$ & 0.0029 & 0.0029 & 0.0023 & 0.0023 & 0.0023\\ 
    & $\sigma \left( \omega_{\mathrm{cdm}} \right)$ & 0.00049 & 0.00024 & 0.00016 & 0.00013 & 7.9e-05\\ 
   $\Lambda$CDM & $\sigma \left( H_0 \right) / [\frac{\mathrm{km}}{\mathrm{s\,Mpc}}]$ & 0.5 & 0.22 & 0.11 & 0.04 & 0.038\\ 
    & $\sigma \left( \ln 10^{10}A_{s } \right)$ & 0.0041 & 0.0038 & 0.0037 & 0.0037 & 0.0018\\ 
   + $M_{\nu}$ & $\sigma \left( n_s \right)$ & 0.0017 & 0.0014 & 0.0006 & 0.00057 & 0.00054\\ 
    & $\sigma \left( \tau_{\mathrm{reio}} \right)$ & 0.0021 & 0.002 & 0.002 & 0.002 & 0.00091\\ 
    & $\sigma \left(M_{\nu}\right) / [eV]$ & 0.041 & 0.019 & 0.012 & 0.0081 & 0.0054\\ 
 \hline\hline   & $\sigma \left( 100*\omega_{\mathrm{b}} \right)$ & 0.0043 & 0.0041 & 0.0039 & 0.0038 & 0.0038\\ 
    & $\sigma \left( \omega_{\mathrm{cdm}} \right)$ & 0.00062 & 0.00052 & 0.00033 & 0.00031 & 0.00029\\ 
   $\Lambda$CDM & $\sigma \left( H_0 \right) / [\frac{\mathrm{km}}{\mathrm{s\,Mpc}}]$ & 0.58 & 0.27 & 0.13 & 0.091 & 0.09\\ 
    & $\sigma \left( \ln 10^{10}A_{s } \right)$ & 0.0043 & 0.0041 & 0.004 & 0.0038 & 0.0022\\ 
   + $M_{\nu}$ & $\sigma \left( n_s \right)$ & 0.0027 & 0.0022 & 0.00061 & 0.00056 & 0.00055\\ 
    & $\sigma \left( \tau_{\mathrm{reio}} \right)$ & 0.0022 & 0.002 & 0.002 & 0.0019 & 0.00092\\ 
    + $N_{\mathrm{eff}}$ & $\sigma \left(M_{\nu}\right) / [eV]$ & 0.04 & 0.02 & 0.013 & 0.0091 & 0.0071\\ 
    & $\sigma \left( N_{\mathrm{eff}} \right)$ & 0.032 & 0.031 & 0.017 & 0.017 & 0.017\\ 
 \hline\hline   & $\sigma \left( 100*\omega_{\mathrm{b}} \right)$ & 0.003 & 0.0028 & 0.0024 & 0.0022 & 0.0023\\ 
    & $\sigma \left( \omega_{\mathrm{cdm}} \right)$ & 0.00053 & 0.00029 & 0.00016 & 0.00014 & 0.0001\\ 
   CDM & $\sigma \left( H_0 \right) / [\frac{\mathrm{km}}{\mathrm{s\,Mpc}}]$ & 4.3 & 0.79 & 0.2 & 0.048 & 0.046\\ 
    & $\sigma \left( \ln 10^{10}A_{s } \right)$ & 0.0042 & 0.0038 & 0.0037 & 0.0037 & 0.0018\\ 
   + $M_{\nu}$ & $\sigma \left( n_s \right)$ & 0.0018 & 0.0014 & 0.0007 & 0.00057 & 0.00054\\ 
    & $\sigma \left( \tau_{\mathrm{reio}} \right)$ & 0.0021 & 0.002 & 0.002 & 0.002 & 0.0009\\ 
   + $w_{0}$ & $\sigma \left(M_{\nu}\right) / [eV]$ & 0.055 & 0.034 & 0.016 & 0.011 & 0.0088\\ 
    & $\sigma \left( w_{0} \right)$ & 0.11 & 0.03 & 0.01 & 0.0021 & 0.0021\\ 
 \hline\hline   & $\sigma \left( 100*\omega_{\mathrm{b}} \right)$ & 0.0029 & 0.0029 & 0.0024 & 0.0023 & 0.0023\\ 
   CDM & $\sigma \left( \omega_{\mathrm{cdm}} \right)$ & 0.00059 & 0.00035 & 0.00018 & 0.00016 & 0.00014\\ 
    & $\sigma \left( H_0 \right) / [\frac{\mathrm{km}}{\mathrm{s\,Mpc}}]$ & 7.1 & 1.6 & 0.25 & 0.056 & 0.053\\ 
   + $M_{\nu}$ & $\sigma \left( \ln 10^{10}A_{s } \right)$ & 0.0042 & 0.0038 & 0.0038 & 0.0037 & 0.0018\\ 
    & $\sigma \left( n_s \right)$ & 0.0018 & 0.0015 & 0.0007 & 0.00058 & 0.00054\\ 
   + $w_{0}$ & $\sigma \left( \tau_{\mathrm{reio}} \right)$ & 0.002 & 0.002 & 0.002 & 0.002 & 0.00092\\ 
    & $\sigma \left(M_{\nu}\right) / [eV]$ & 0.055 & 0.048 & 0.023 & 0.017 & 0.015\\ 
   + $w_{a}$ & $\sigma \left( w_{0} \right)$ & 0.28 & 0.17 & 0.021 & 0.0022 & 0.0021\\ 
    & $\sigma \left( w_{a} \right)$ & 1.8 & 0.41 & 0.067 & 0.016 & 0.016\\ 
 \hline\end{tabular} 
 \endgroup}